\documentclass[a4paper,twocolumn,8pt,accepted=2021-12-15]{quantumarticle}
\pdfoutput=1
\usepackage[utf8]{inputenc}
\usepackage[english]{babel}
\usepackage[T1]{fontenc}

\usepackage{hyperref}

\usepackage{lipsum}

\usepackage{graphicx}
\usepackage{dcolumn}
\usepackage{bm}

\usepackage{multirow}

\usepackage{amsmath,amssymb,amsfonts}
\usepackage{graphicx}
\usepackage{color}
\usepackage{appendix}
\usepackage{amsthm}
\usepackage{tikz}

\usepackage{mleftright}
\usepackage{url}

\usepackage{hyperref}

\usepackage{tikz}
\usetikzlibrary{shapes,snakes,positioning,automata,arrows.meta,calc,decorations.markings,math,arrows.meta}
\tikzstyle{vertex}=[circle, draw, inner sep=0pt, minimum size=6pt]
\usepackage{xcolor}
\usepackage{subfig}

\usepackage[export]{adjustbox}

\usepackage{hyperref}

\usepackage[english]{babel}
\usepackage[utf8]{inputenc}

\usepackage{cite}

\usepackage{geometry}
\newtheorem{theorem}{Theorem}

\newtheorem{lemma}{Lemma}
\newtheorem{definition}{Definition}

\newcommand{\nn}{\nonumber}

\def\a{\alpha}
\def\b{\beta}

\def\p{\psi}
\def\w{\omega}

\def\P{\Psi}

\def\<{\langle}
\def\>{\rangle}

\def\mH{{\mathcal{H}}}

\def\ha{{\hat{a}}}
\def\hb{{\hat{b}}}

\def\vn{{\vec{n}}}

\def\dc{{\overline{G}_d }}

\begin{document}

\title{Graph picture of linear quantum networks and entanglement}

	\author{Seungbeom Chin }
	\email{sbthesy@gmail.com}
	\affiliation{Department of Electrical and Computer Engineering, Sungkyunkwan University, Suwon 16419, Korea \\ International Centre for Theory of Quantum Technologies, University of Gd{\'a}nsk, 80-308, Gd{\'a}nsk, Poland}

	\author{Yong-Su Kim }
	\affiliation{Center for Quantum Information, Korea Institute of Science and Technology (KIST), Seoul, 02792, Korea \\
		Division of Nano $\&$ Information Technology, KIST School, Korea University of Science and Technology, Seoul 02792, Korea}
	
	\author{Sangmin Lee}
	\email{sangmin@snu.ac.kr}
	\affiliation{ College of Liberal Studies, Seoul National University, Seoul 08826, Korea}

\maketitle

\begin{abstract}
		The indistinguishability of quantum particles is widely used as a resource for the generation of entanglement. Linear quantum networks (LQNs), in which identical particles linearly evolve to arrive at multimode detectors, exploit the indistinguishability to generate various multipartite entangled states by the proper control of transformation operators. However, it is challenging to devise a suitable LQN that carries a specific entangled state or compute the possible entangled state in a given LQN as the particle and mode number increase. This research presents a mapping process of arbitrary LQNs to graphs, which provides a powerful tool for analyzing and designing LQNs to generate multipartite entanglement. We also introduce the perfect matching diagram (PM diagram), which is a refined directed graph that includes all the essential information on the entanglement generation by an LQN. The PM diagram furnishes rigorous criteria for the entanglement of an LQN and solid guidelines for designing suitable LQNs for the genuine entanglement. Based on the structure of PM diagrams, we compose LQNs for fundamental $N$-partite genuinely entangled states.
\end{abstract}

	\section{Introduction} \label{intro}

	Entanglement is an essential property of multipartite quantum systems that works as a resource for several kinds of quantum information processing~\cite{horodecki2009quantum}. Among diverse methods to create entangled multipartite systems, the indistinguishability of quantum particles is frequently employed for obtaining such systems. As a well-known example, Hong-Ou-Mandel interference~\cite{hong1987measurement} showed that if two fully-indistinguishable photons are scattered so that their paths completely overlap, the two-photon state is entangled at two detectors (a $N00N$ state with $N=2$). Based on a similar intuition that the path ambiguity of indistinguishable particles can generate entanglement, various protocols have been suggested for the entanglement of identical  particles~\cite{tichy2013entanglement,killoran2014extracting,krenn2017entanglement, paunkovic2004role}. Recently, there have been quantitative studies on the relation among  particle indistinguishability, spatial coherence, and bipartite entanglement~\cite{franco2018indistinguishability, chin2019entanglement, nosrati2020robust, barros:20, chin2021taming}. In particular, the amount of entanglement is experimentally verified to be a monotonically increasing function of both indistinguishability and spatial coherence~\cite{barros:20}.
	
	While the above works mainly focus on bipartite systems, another direction for exploiting the particle indistinguishability is the construction of general $N$-partite genuinely entangled states.  Tripartite GHZ state generation with particles from independents sources was devised in Refs.~\cite{yurke1992einstein,yurke1992bell}, whose method was generalized in Refs.~\cite{zukowski1993event,zukowski1995entangling,zeilinger1997three}. More recently, Ref.~\cite{blasiak2019entangling} designed linear multi-rail encoding networks for three photons to obtain tripartite GHZ and W states and Refs.~\cite{bellomo2017n,kim2020efficient} suggested protocols for generating the $N$-partite W-state with spatially overlapped $N$ identical particles and an ancilla. 
	
	\begin{widetext}
		\begin{center}
	\begin{figure}
		\subfloat[An $N=4$ LQN.  Each particle, initially at $\p_i$ (spatial wavefunction, $i=1,2,3,4$), evolves by the linear transformation operator $T$ and arrives at the detectors $X_i$ $(i=1,2,3,4)$. The blue (red) color denotes the internal state $\uparrow$ ($\downarrow$).]{
			\begin{tikzpicture}[baseline]
				\draw[fill=gray!40!white, very thick, draw=gray!20!white ] (0,-.2) -- (1,-.2) arc(0:-180:.5) --cycle;
				\draw[fill=gray!40!white, very thick, draw=gray!20!white] (1.5,-.2) -- (2.5,-.2) arc(0:-180:.5) --cycle;
				\draw[fill=gray!40!white, very thick, draw=gray!20!white] (3.1,-.2) -- (4.1,-.2) arc(0:-180:.5) --cycle;      
				\draw[fill=gray!40!white, very thick, draw=gray!20!white] (4.6,-.2) -- (5.6,-.2) arc(0:-180:.5) --cycle;                
				\draw (0.5,-0.4) node {$X_1$} ; 
				\draw (2,-0.4) node {$X_2$} ;
				\draw (3.6,-0.4) node {$X_3$} ;
				\draw (5.1,-0.4) node {$X_4$} ;	
				\draw (0.5,2.5) node {$(\p_1,\downarrow)$} ;  
				\draw (2,2.5) node {$(\p_2,\downarrow)$} ;    
				\draw (3.6,2.5) node {$(\p_{3},\uparrow)$} ;  
				\draw (5.1,2.5) node {$(\p_{4},\uparrow)$} ; 
				\draw[fill=red!60!white, thick, draw= red!20!white] (0.5,2) circle (0.2);   
				\draw[fill=red!60!white, thick, draw= red!20!white] (2,2) circle (0.2);   	
				\draw[fill=blue!60!white, thick, draw= blue!20!white] (3.6,2) circle (0.2); 
				\draw[fill=blue!60!white, thick, draw= blue!20!white] (5.1,2) circle (0.2); 
				\draw[red!60!white,thick, -{Stealth[scale=1.2]}] (0.4,1.7) -- (0.4, 0);				
				\draw[red!60!white,thick, -{Stealth[scale=1.2]}] (0.6,1.7) -- (3.4, 0);
				\draw[red!60!white,thick, -{Stealth[scale=1.2]}] (0.7,1.7) -- (4.9, 0);				
				
				\draw[red!60!white,thick, -{Stealth[scale=1.2]}] (2,1.7) -- (2, 0);		
				\draw[red!60!white,thick, -{Stealth[scale=1.2]}] (2.1,1.7) -- (3.5, 0);
				\draw[red!60!white,thick, -{Stealth[scale=1.2]}] (2.2,1.7) -- (5, 0);	
				\draw[blue!60!white,thick,-{Stealth[scale=1.2]}] (3.4,1.7) -- (0.6, 0);				
				\draw[blue!60!white,thick, -{Stealth[scale=1.2]}] (3.5,1.7) -- (2.1, 0);		
				\draw[blue!60!white,thick, -{Stealth[scale=1.2]}] (3.6,1.7) -- (3.6, 0);
				\draw[blue!60!white,thick,-{Stealth[scale=1.2]}] (4.9,1.7) -- (0.7, 0);				
				\draw[blue!60!white, thick, -{Stealth[scale=1.2]}] (5,1.7) -- (2.2, 0);		
				\draw[blue!60!white, thick, -{Stealth[scale=1.2]}] (5.2,1.7) -- (5.2, 0);			
				\draw[rounded corners=4, dashed, thick] (0.1,0.4) rectangle (5.5,1.5);
				\draw (-0.2,1) node {$T$};		 			  
		\end{tikzpicture}}\qquad 
		\subfloat[The $G_{bb}$ and $G_d$ of the above LQN. From the structure of the graphs, we can directly read that the LQN carries a Dicke state. For a complete analysis of the entanglement in the graphs, see Section~\ref{applications}.]{
			\begin{tikzpicture}[baseline]
				\node[circle,draw] (1) at (4.5,0.4) {$w_3 $};
				\node[circle,draw] (2) at (6.7,0.4) {$w_4 $};
				\node[circle,draw] (3) at (6.7,2.6) {$w_2$};
				\node[circle,draw] (4) at (4.5,2.6) {$w_1$};
				\path[every loop/.style={min distance=10mm,in=-180,out=-90,looseness=5,},line width = 0.5pt,->,color=blue] (1) edge[loop above]   (1);
				\path[every loop/.style={min distance=10mm,in=-90,out=0,looseness=5},line width = 0.5pt,->,color=blue] (2) edge [loop above]   (2);
				\path[every loop/.style={min distance=10mm,in=0,out=90,looseness=5,color=red},line width = 0.5pt,->] (3) edge [loop above]   (3);	
				\path[every loop/.style={min distance=10mm,in=90,out=180,looseness=5,color=red},line width = 0.5pt,->] (4) edge [loop above]   (4);
				%
				\path[line width = 0.8pt,->,color=blue] (1) edge[bend right=15]  node[near start] {$ $ } (3);
				\path[line width = 0.8pt,->,color=blue] (1) edge[bend right=15]  node[near start] {$ $ } (4);
				%
				\path[line width = 0.8pt,->,color=blue] (2) edge[bend right=15]  node[near start] {$ $ } (3);
				\path[line width = 0.8pt,->,color=blue] (2) edge[bend right=15]  node[near start] {$ $ } (4);
				\path[line width = 0.8pt,color=red,->] (3) edge[bend right=15]  node[near start] {$ $ } (2); 
				\path[line width = 0.8pt,color=red,->] (3) edge[bend right=15]  node[near start] {$ $ } (1);	
				\path[line width = 0.8pt,color=red,->] (4) edge[bend right=15]  node[near start] {$ $ } (1); 
				\path[line width = 0.8pt,color=red,->] (4) edge[bend right=15]  node[near start] {$ $ } (2);	
				%
				%
				\node[vertex] (1) at (0,3) {$1 $} ;
				\node[vertex] (2) at (0,2) {$2 $};
				\node[vertex] (3) at (0,1) {$3 $};
				\node[vertex] (4) at (0,0) {$4 $};	
				\node[circle,draw] (X) at (2.5,3) {$X_1$};
				\node[circle,draw] (Y) at (2.5,2) {$X_2$};
				\node[circle,draw] (Z) at (2.5,1) {$X_3$};
				\node[circle,draw] (W) at (2.5,0) {$X_4$};	
				\path[line width = 0.8pt,color=red] (1) edge  (X);
				\path[line width = 0.8pt,color=red ] (1) edge  (Z);
				\path[line width = 0.8pt,color=red ] (1) edge (W);
				\path[line width = 0.8pt,color=red ] (2) edge (Y);
				\path[line width = 0.8pt,color=red] (2) edge (Z);	
				\path[line width = 0.8pt,color=red] (2) edge  (W);
				\path[line width = 0.8pt,color=blue ] (3) edge  (X);
				\path[line width = 0.8pt,color=blue ] (3) edge (Y);
				\path[line width = 0.8pt,color=blue ] (3) edge (Z);
				\path[line width = 0.8pt,color=blue ] (4) edge (X);
				\path[line width = 0.8pt,color=blue] (4) edge  (Y);
				\path[line width = 0.8pt,color=blue ] (4) edge  (W);
		\end{tikzpicture}}
		\caption{A linear quantum network (LQN) with $N=M=4$ and its corresponding graphs.}
		\label{linearq}
	\end{figure}
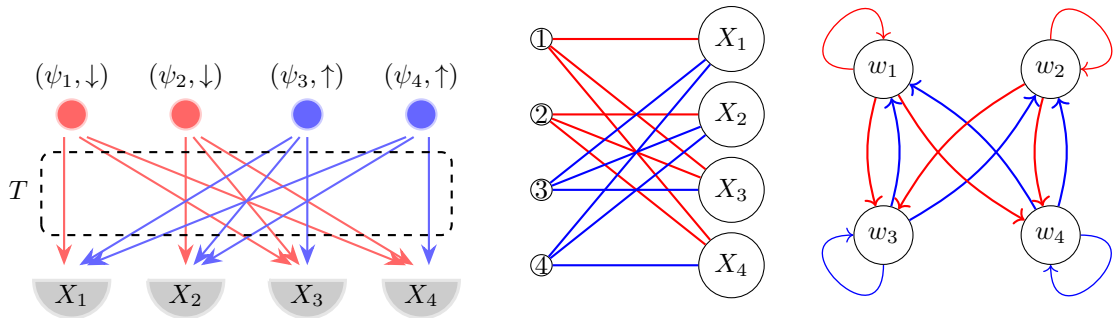
		\end{center}
	\end{widetext}

	Among the various approaches to obtain entanglement with identical particle, a powerful method that is theoretically simple and experimentally scalable is to build \emph{linear quantum networks} (LQN), in which each identical particle state is expressed as a local field that evolves linearly in space and time. An LQN consists of three elements (Fig.~\ref{linearq}.a): $N$ identical particles with a $d$-dimensional internal degree of freedom, a passive transformation operator $T$ that transforms the initial particle states so that the final particle states are the linear combinations of the initial states, and $M$ detectors that are distinctively located to each other. The $N$ particles that evolve according to $T$ are observed at $M$ detectors. We can generate a variety of multipartite entangled states by controlling $T$. Indeed, the spatial overlap (or coherence) that is considered an essential factor for the entanglement in Refs.~\cite{franco2018indistinguishability, chin2019entanglement,barros:20} is a special form of $T$.   
	
	For an LQN, the ratio of $N$ and $M$ is a crucial factor for the non-classicality of quantum systems. For example, when $N\gg M$ with bosons, the system exhibits the Bose-Einstein condensation that can be simulated efficiently with a semiclassical approach~\cite{urbina2016multi}.
	On the other hand, when $N \ll M$, ideally indistinguishable bosons ($d=1$) scattered by a linear unitary operator constitute the boson sampling setup~\cite{aaronson2011computational}. Boson sampling is a feasible implementation whose quantum advantage has been experimentally demonstrated~\cite{broome2013photonic,spring2013boson,wang2019boson}.

	In this work, we focus on another intriguing case where $N=M$ and each identical particle has a two-dimensional internal degree of freedom, i.e., $d=2$ (see Fig.~\ref{linearq}.a for an $N=M=4$ example). At the measurement level, we postselect the states for each detector to observe only one particle, i.e., \emph{no particle-bunching states}. Then each detector becomes a subsystem whose state is determined by the internal state of the particle that is detected.
		Thus, the two internal states begin to play the role of qubits. Many schemes of LQNs for generating entanglement is based on this setup~\cite{krenn2017entanglement,bellomo2017n,franco2018indistinguishability,kim2020efficient,blasiak2019entangling}. It was also shown  that such a postselection is safe from the selection bias i.e., the correlation in the final state is not fully created by the postselection~\cite{blasiak2020safe}. It is worth noting that we consider the cases where the identical particles are ideal, i.e., fully indistinguishable. 
	
		On the other hand, a critical limitation of the previous research on generating entanglement in this setup is \emph{the lack of manifest insights in the relations between the structure of a LQN (particle paths and internal state distributions) and the entanglement in it.} In other words, one cannot verify the entanglement in the final postselected state of a given LQN until completing all the computation processes. For this reason, the research on the entanglement of LQN have remained in the case-by-case approach. Our main goal is to overcome the limitation with a graph picture. By introducing a systematic strategy to grasp the quantitative relation between LQN structures and entanglement, our graph picture provides a powerful insight for analyzing the entanglement of identical particles carried by LQNs.
	
	The advantage of associating graph theory with quantum information processing has been validated in many works: contextuality in the graph and hypergraph frameworks~\cite{cabello2014graph,acin2015combinatorial}, quantum graph states~\cite{dur2003multiparticle,hein2004multiparty, van2005local,hein2006entanglement}, Gaussian boson sampling exploited to solve the computational complexity problems in graph theory~\cite{bradler2018gaussian,arrazola2018using,bradler2021graph}, and  the linkage of undirected graphs to optical setups involving multiple crystals~\cite{krenn2017quantum,gu2019quantum,gu2019quantum2,gu2020quantum}
	Here, we show that \emph{the graph picture of LQNs is also beneficial for computing the entanglement of no-bunching states in LQNs}. Moreover, by showing that the graph structure is closely related to the entanglement property of the corresponding LQNs, \emph{we can reversely propose a protocol for designing optimal LQNs that generate specific $N$-partite entangled states.} 
	
	
	The result of this work is divided into two parts: 
	The first part explains how LQNs of our interest can be mapped into graphs, and exploits graph-theoretical techniques for computing the entanglement of LQNs.   
	We achieve a graph picture of an LQN by inserting all the indispensable quantities in the LQN into the weighted and colored adjacency matrices $A$ and $C$, which combine to draw graphs of colored weighted edges. By slightly varying the mapping relations, an LQN can be represented as two types of graphs, a \emph{balanced bipartite graph} $G_{bb}$ and a \emph{directed graph} $G_d$ (the mathematical definitions of the graph theory are summarized in Appendix~\ref{glossary}). Fig.~\ref{linearq}.b presents the graph picture of an $N=4$ example.
	We show how the combined use of the two representations provides a convenient protocol to compute the final no-bunching state that we postselect.
	
	The second part introduces the concept of \emph{perfect matching (PM) diagram} $\dc$, which is a refined $G_d$-representation of LQNs that includes the essential information of LQNs. A $\dc$ corresponds to an LQN that only contains the particle paths that contribute to the postselected no-bunching states. Hence, we can analyze the entanglement property of an LQN by examining the $\dc$ structure. Moreover, the $\dc$-representation provides strong guidelines to build LQNs that carry specifically entangled no-bunching states. Indeed, we suggest LQNs that generates the $N$-partite GHZ class, W class, and Dicke states designed from the $\dc$ structure. 
	
	Therefore, we can state that \emph{the graph picture of LQNs enables a schematic approach to the two central procedures for the entanglement of LQNs}, i.e., designing an appropriately entangled LQN and extracting no-bunching states in the LQN.

	\section{Computing LQN states with graphs}\label{LQNstoGraphs}
	
	Here, we provide the mathematical formulation of LQNs and then explain how an LQN is mapped to two forms of graphs, $G_{bb}$ and $G_{d}$. 
	We are interested in the postselection of no-bunching states and the evaluation of the entanglement carried by the states. Thus, 
	an essential part of quantifying the entanglement is to compute the no-bunching states. Using the fact that such states correspond to \emph{perfect matchings} (PMs, see Appendix~\ref{glossary}) in the $G_{bb}$-representation, we propose a simple and intuitive protocol for obtaining the total no-bunching state in an LQN. 
	
	\subsection{Linear quantum network (LQN)}
	
	A quantum network of $N$ particles, which evolve in spacetime and arrive at $M$ spatially distinctive detectors, is \emph{linear} if the initial state of each particle is linearly related to the final detector states, and vice versa. As we have mentioned in the introduction, we study the LQNs of $N=M$ and two-dimensional internal state ($d=2$) with the computational basis $\{\uparrow,\downarrow \}$.
	
	We denote the initial state of an $a$th particle as $\P_a = (\p_a, s_a)$ ($\p_a$: the particle spatial wavefunction, $s_a$: the internal state, and $a=1,\cdots, N$), and the detector spatial wavefunctions as $\phi_j$ ($j=1,\cdots, N$). The wavefunctions are set to be orthonormal without loss of generality. 
	Then, the linear evolution of a particle in $\P_a$ to the $j$th detector is expressed as 
	\begin{align}\label{linear_single}
		|\P_a\>=	|\p_a,s_a\> \to  \sum_{j=1}^N T_{aj}|\phi_j, r^a_j\>, \qquad  (T_{aj} \in \mathbb{C})  
	\end{align} where the transformation amplitude $T_{aj}$ is normalized as $\sum_{j}|T_{aj}|^2 =1$ so that the particle states evolve probabilistically, and $r^a_j$ is the internal state of a particle that leaves $\p_a$ for $\phi_j$~\footnote{Eq.~\eqref{linear_single} is a simplified form of the general transformation relation $|\P_a\>=	|(\p_a,s_a)\> = \sum_{j=1}^N\sum_{r_j=0}^{d-1} T_{a,jr_j}|(\phi_j, r_j)\>$ $(T_{a,jr_j} \in \mathbb{C})$ where the complex number $T_{a,jr_j}$ is normalized as $\sum_{j,r_j}|T_{a,jr_j}|^2 =1$. A detailed derivation is given in Appendix~\ref{lineartransformation}. }. Note that we can set $r_j^a$ to be different from $s_a$, which shows that $r_j^a$ is a part of the linear transformation. This kind of internal state transformations can be realized with, e.g., polarizing beamsplitters (PBS) in linear optics~\cite{barros:20,lee2021entangling} or Stern-Gerlach apparatus in spin-particle systems.
	The above equation is equivalently expressed with the field operators of the second quantization language as    
	\begin{align}\label{transf_2ql}
		\ha_{as_a}^\dagger \to
		\sum_{j=1}^N T_{aj}\hat{b}_{jr^a_j}^\dagger,
	\end{align} where $\ha^\dagger_{as_a}|vac\> \equiv |\p_a,s_a\>$ and $\hb^\dagger_{jr^a_j}|vac\> \equiv |\phi_j,r^a_j\>$.
	
	Then the total $N$-particle state transformation is given by
	\begin{align}\label{N_transf}
		\prod_{a=1}^N\ha_{as_a}^\dagger|vac\>\to   \prod_{a=1}^N \Big(\sum_{j=1}^N T_{aj}\hat{b}_{j r^a_{j} }^\dagger\Big)|vac\>.	
	\end{align}
	
	At the detector level, we postselect only the \emph{no-bunching states}, i.e., the cases when each detector observes only one particle. Then the superposition of such states determine the entanglement observed at the output detectors. 

	\begin{figure}
		\centering
		\begin{tikzpicture}
			\draw[fill=gray!40!white, very thick, draw=gray!20!white ] (3,-0.5) -- (3,0.5) arc(90:-90:.5) --cycle;
			\draw[fill=gray!40!white, very thick, draw=gray!20!white] (3,1) -- (3,2) arc(90:-90:.5) --cycle;
			\draw (3.25,1.5) node {$X_1$} ; 
			\draw (3.25,0) node {$X_2$} ;
			
			
			\draw[fill=red!40!blue, thick, draw= black!20!white] (0,0) circle (0.2);   
			\draw[fill=red!40!blue, thick, draw= black!20!white] (0,1.5) circle (0.2);   	
			
			\draw[red!60!white,thick, -{Stealth[scale=1.2]}] (0.3,-.1) -- (2.9, -.1);				
			\draw[blue!60!white,thick, -{Stealth[scale=1.2]}] (0.3,.1) -- (2.9,1.4);		
			
			\draw[red!60!white,thick, -{Stealth[scale=1.2]}] (0.3,1.6) -- (2.9, 1.6);				
			\draw[blue!60!white,thick, -{Stealth[scale=1.2]}] (0.3,1.4) -- (2.9, 0.1);		
			
			\draw (-1,0) node {$(\p_2,s_2)$};
			\draw (-1,1.5) node {$(\p_1,s_1)$};
		\end{tikzpicture}
		\caption{An $N=2$ LQN example. An entangled no-bunching state can be detected at the spatially separated two detectors.}
		\label{N=2_BS}
	\end{figure}
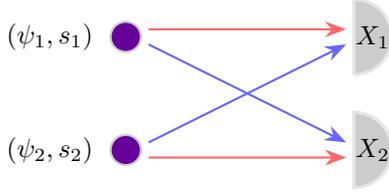

	As a simple example, suppose that two identical particles at two separated places $\ha^\dagger_{1s_1}$ and $\ha^\dagger_{2s_2}$ ($|s_1\> = \a_1|\uparrow\> + \b_1|\downarrow\>$ and $|s_2\> = \a_2|\uparrow\> + \b_2|\downarrow\>$) transform as
	\begin{align}
		\begin{split}
			\ha^\dagger_{1s_1} &\to \a_1\hb^\dagger_{1\uparrow} +\b_1\hb^\dagger_{2\downarrow},
			\\
			\ha^\dagger_{2s_2} &\to  \a_2\hb^\dagger_{1\downarrow} +\b_2\hb^\dagger_{2\uparrow}. 
		\end{split}
	\end{align} 
	with $|\a_1|^2 + |\b_1|^2 = |\a_2|^2 + |\b_2|^2 = 1$ (see Fig.~\ref{N=2_BS}). 
 
	In the form of Eq.~\eqref{N_transf}, we set
	\begin{align}\label{N=2_Tr}
		&(T_{11},T_{12},T_{21},T_{22}) = (\a_1,\b_1,\a_2,\b_2), \nn \\ &(r^1_1,r^1_2,r^2_1,r^2_2) = (\uparrow,\downarrow,\downarrow,\uparrow).
	\end{align}  Then the state transformation is given by 
	\begin{align}\label{N=2}
		\ha^\dagger_{1s_1}\ha^\dagger_{2s_2}|vac\> &\to (\a_1\hb^\dagger_{1\uparrow} + \b_1\hb^\dagger_{2\downarrow}) (\a_2\hb^\dagger_{1\downarrow} + \b_2\hb^\dagger_{2\uparrow})|vac\> \nn \\
		&=(\a_1\a_2\hb^\dagger_{1\uparrow}\hb^\dagger_{1\downarrow} +  \a_1\b_2\hb^\dagger_{1\uparrow}\hb^\dagger_{2\uparrow} \nn \\
		&\quad  + \b_1\a_2\hb^\dagger_{2\downarrow}\hb^\dagger_{1\downarrow}+ \b_1\b_2\hb^\dagger_{2\downarrow}\hb^\dagger_{2\uparrow})|vac\>.
	\end{align}
	By postselecting no-bunching states, the (unnormalized) final state to be measured is given by
	\begin{align}\label{N=2_fin}
		|\P_\mathrm{fin}\>&=(\a_1\b_2\hb^\dagger_{1\uparrow}\hb^\dagger_{2\uparrow}+ \b_1\a_2\hb^\dagger_{2\downarrow}\hb^\dagger_{1\downarrow})|vac\> \nn \\
		&\equiv \a_1\b_2|\uparrow_1,\uparrow_2\> \pm \b_1\a_2|\downarrow_1,\downarrow_2\>. 
	\end{align} where $+$ for bosons and $-$ for fermions in the last line. This is an entangled state of mode $X_1$ and $X_2$ with concurrence  $E_c(|\P_\mathrm{fin}\>)=4|\a_1\a_2\b_1\b_2|$~\cite{franco2018indistinguishability,chin2019entanglement,barros:20}. Note that $E_c=0$ when either of the particles arrives at one detector with probability 1 (no spatial coherence).

	\subsection{Graph picture of LQN}\label{LQNtoG}
	
	While the above $N=2$ example is relatively easy to treat, it becomes quickly complicated as $N$ increases to evaluate the entanglement carried by no-bunching states in an LQN. On the other hand, we can map LQNs into graphs, in which the computation usually becomes much easier. Here, we formalize  the mapping of LQNs into graphs.
	
	The graph picture of an LQN can be achieved by inserting all the indispensable variables of LQN in Eq.~\eqref{N_transf} into the weighted adjacency matrix $A$ and the colored adjacency matrix $C$, which combine to draw a graph with colored weighted edges.
	Eq.~\eqref{N_transf} denotes that a particle in ($\p_a,s_a$) arrives at the the $j$th detector with probability $|T_{aj}|^2$ and  internal state $r^a_j$, which is set to be $\uparrow$ or $\downarrow$~\footnote{In the most general sense, we can consider a transformation $T$ so that $r_j^a$ becomes a superposition of $\uparrow$ and $\downarrow$, which we put off to our future research. Such an LQN includes, e.g., partially polarizing beam splitter (PPBS) in linear optics~\cite{kim2012protecting,langford2005demonstration,kiesel2005linear,okamoto2005demonstration}.}. Note that $s_a$ becomes irrelevant to the final state since it is replaced by $r^a_j$ in the transformation process. Thus, $T_{aj}$ and $r^a_j$ are the quantities that we must keep through the mapping. 
	
	As a result, if we  insert $T_{aj}$ into the elements of $A$ and $r^a_j$ into those of $C$ as
	\begin{align}\label{reduction}
		A_{aj} = T_{aj}\quad (\in \mathbb{C}), \qquad 
		C_{aj} = r^a_j \quad (\in \mathbb{Z}_2), 
	\end{align} the graph determined by $A$ and $C$ fully contains the relevant information on the LQN.   
	Note that $A$ and $C$ always have the same number of non-zero entries.

	Considering the relation of adjacency matrices with graphs (see Appendix~\ref{glossary}), the mapping relations Eq.~\eqref{reduction} imply that the particle index $a$ and the detector index $j$ in LQN become the vertex indices, and the internal state $r^a_j$ becomes the color of an edge connecting vertices $a$ and $j$ in the graph picture. We set the edge color blue (B) for $r^a_j=\uparrow$ and red (R) for $r^a_j=\downarrow$.

	In graph theory, it is well-known that an $(N\times N)$-adjacency matrix can construct two types of graphs, i.e., a \emph{directed graph} $G_d$ and an \emph{undirected balanced bigraph} $G_{bb}$, according to how we define the relation of indices $a$ and $j$ with vertices (see, e.g., Ref.~\cite{brualdi1980bigraphs}). 
	
	First, we can consider both $a$ and $j$ as the indices of a $N$-vertex set $W =\{w_1,w_2,\cdots,w_N\}$. Then, $A_{aj}$ ($C_{aj}$) indicate the weight (color) of an edge from $w_a$ to $w_j$. Such $(N\times N)$-matrices, denoted as $A_d$ and $C_d$, become the adjacency matrices of a directed graph $G_d=(W,F)$. In this $G_{d}$-representation, one particle and one detector are mapped to the same vertex in $W$.
	
	On the other hand, we can also consider $a$ and $j$ as indices of two different vertex sets,  i.e., $a$ as the index of a vertex set $U =\{u_1,u_2,\cdots,u_N\}$ and  $j$ as that of another set $V =\{v_1,v_2,\cdots,v_N\}$. 
	Then a balanced bigdraph $G_{bb}=(U\cup V, E)$ is constructed by setting the adjacency matrices $A_{bb}$ and $C_{bb}$ as symmetric forms 
	\begin{align}\label{G_bb}
		A_{bb} = \begin{pmatrix}
			0 & A_{d}\\
			(A_{d})^T & 0 
		\end{pmatrix}, \quad
		C_{bb} = \begin{pmatrix}
			0 & C_{d}\\
			(C_{d})^T & 0 
		\end{pmatrix},
	\end{align} where $(A_d)^T$ and $(C_{d})^T$ are the transposes of $A_d$ and $C_d$.  
	From the fact that particles and detectors are mapped distinctively, the $G_{bb}$-representation depicts an LQN in a very intuitive manner. However, the $G_d$-representation is also advantageous for analysing some crucial hidden structures of the LQN, which we will discuss in Section~\ref{G_d}. 
	
	An important feature of $A_d$ and $C_d$ from an LQN is that they are \emph{symmetric (antisymmetric) under the exchange of rows according to the exchange symmetry (antisymmetry) of bosons (fermions) in the LQN.} Therefore, if two $G_{bb}$ become identical by any exchange of vertices in $U$, we consider them an equivalent graph that corresponds to the same LQN.
	
	To elucidate the mapping process of an LQN into graphs, we see how the $N=2$ example in Fig.~\ref{N=2_BS} is mapped to graphs. By substituting Eq.~\eqref{N=2_Tr} into Eq.~\eqref{reduction}, $A_{d}$ and $C_d$ in the $G_d$-representation are given by
	\begin{align}
		&A_d= \begin{pmatrix}
			\a_1 & \b_1 \\
			\a_2 & \b_{2} \\
		\end{pmatrix}\cong \pm \begin{pmatrix}
			\a_2 & \b_2\\
			\a_1 & \b_{1} \\
		\end{pmatrix}
		,\nn \\
		&C_d = 
		\begin{pmatrix}
			B & B \\
			R & R \\
		\end{pmatrix}
		\cong  \pm \begin{pmatrix}
			R & R \\
			B & B \\
		\end{pmatrix} \nn\\
		& (\textrm{$+$ for bosons and $-$ for fermions})
	\end{align} ($\cong$ means two adjacency matrices are physically identical by the exchange symmetry),    
	which draws a $G_d$,
	\begin{align}
		&\begin{tikzpicture}[baseline={([yshift=-.5ex]current bounding box.center)}]
			\node[circle,draw,minimum size=0.5cm] (XX) at (0,0) {$w_1$};
			\node[circle,draw,minimum size=0.5cm] (YY) at (2,0) {$w_2$};
			\path[every loop/.style={min distance=10mm,in=135,out=225,looseness=5},line width = 0.5pt,->,color=blue] (XX) edge [loop above] node {$\a_1$} (XX);
			\path[every loop/.style={min distance=10mm,in=45,out=-45,looseness=5},line width = 0.5pt,->,color=red] (YY) edge [loop above] node{$\b_2$} (YY);		
			\path[line width = 0.8pt, ->,color=blue] (XX) edge[bend left]   node[ near start] {$\b_1$ } (YY);
			\path[line width = 0.8pt,color=red, -> ] (YY) edge[bend left]  node[near start] {$\a_2$ } (XX);
		\end{tikzpicture}
		\nn \\
		\cong &\pm 
		\begin{tikzpicture}[baseline={([yshift=-.5ex]current bounding box.center)}]
			\node[circle,draw,minimum size=0.5cm] (XX) at (5.5,0) {$w_1$};
			\node[circle,draw,minimum size=0.5cm] (YY) at (7.5,0) {$w_2$};
			\path[every loop/.style={min distance=10mm,in=135,out=225,looseness=5},line width = 0.5pt,->,color=red] (XX) edge [loop above] node {$\a_2$} (XX);
			\path[every loop/.style={min distance=10mm,in=45,out=-45,looseness=5},line width = 0.5pt,->,color=blue] (YY) edge [loop above] node{$\b_1$} (YY);		
			\path[line width = 0.8pt, ->,color=red] (XX) edge[bend left]   node[ near start] {$\b_2$ } (YY);
			\path[line width = 0.8pt,color=blue, -> ] (YY) edge[bend left]  node[near start] {$\a_1$ } (XX);
		\end{tikzpicture}.
	\end{align}
	For the same $N=2$ LQN, $A_{bb}$ and $C_{bb}$ in the $G_{bb}$-representation are given by
	\begin{align}\label{N=2_bipartite1}
		A_{bb}
		&= \begin{pmatrix}
			0&0&\a_1 & \b_1 \\
			0&0&\a_2 & \b_{2} \\
			\a_1 & \a_2 &0&0 \\
			\b_1 &\b_2 &0&0 \\ 
		\end{pmatrix} \cong 
		\pm \begin{pmatrix}
			0&0&\a_2 & \b_2\\
			0&0&\a_1 & \b_{1} \\
			\a_2 & \a_1 &0&0 \\
			\b_2 &\b_1 &0&0 \\ 
		\end{pmatrix}, \nn \\
			C_{bb}
		&= \begin{pmatrix}
			0&0& B & B \\
			0&0& R & R \\
			B & R &0&0 \\
			B & R &0&0 \\ 
		\end{pmatrix} \cong 
		\pm \begin{pmatrix}
			0&0&R & R\\
			0&0&B & B \\
			R & B &0&0 \\
			R & B &0&0 \\ 
		\end{pmatrix},
	\end{align}
	which draws a $G_{bb}$,
	\begin{align}\label{N=2_bipartite}
		\begin{tikzpicture}[baseline={([yshift=-.5ex]current bounding box.center)}]
			\node[vertex] (1) at (0,2.5) {$1$};
			\node[vertex] (2) at (0,1) {$2$};  
			\node[circle,draw,minimum size=0.8cm] (X) at (2.5,2.5) {$X_1$};
			\node[circle,draw,minimum size=0.8cm] (Y) at (2.5,1) {$X_2$};
			\path[line width = 0.8pt, color=blue ] (1) edge   node[near start] {$\a_1$ } (X);
			\path[line width = 0.8pt, color=red] (2) edge  node[near start] {$\b_2$ } (Y);
			\path[line width = 0.8pt, color=blue] (1) edge   node[near start] {$\b_1$ } (Y);
			\path[line width = 0.8pt,color=red] (2) edge  node[near start] {$\a_2$ } (X);
		\end{tikzpicture}
		\cong  \pm
		\begin{tikzpicture}[baseline={([yshift=-.5ex]current bounding box.center)}]
			\node[vertex] (1) at (0,2.5) {$1$};
			\node[vertex] (2) at (0,1) {$2$};  
			\node[circle,draw,minimum size=0.8cm] (X) at (2.5,2.5) {$X_1$};
			\node[circle,draw,minimum size=0.8cm] (Y) at (2.5,1) {$X_2$};
			\path[line width = 0.8pt, color=red ] (1) edge   node[near start] {$\a_2$ } (X);
			\path[line width = 0.8pt, color=blue] (2) edge  node[near start] {$\b_1$ } (Y);
			\path[line width = 0.8pt, color=red] (1) edge   node[near start] {$\b_2$ } (Y);
			\path[line width = 0.8pt,color=blue] (2) edge  node[near start] {$\a_1$ } (X);
		\end{tikzpicture}.
	\end{align} 
	In the above $G_{bb}$, we used $a$ ($=1,2,\cdots, N$) for the vertex index of $U$ and $X_j$ ($j=1,2,\cdots, N$) for that of $V$, which will be the standard notation from now on.
	By comparing Eq.~\eqref{N=2_bipartite} with Fig.~\ref{N=2_BS}, we see that the $G_{bb}$-representation straightforwardly describes the LQN structure.  Two vertices with indices 1 and 2 denote two identical particles. The edges correspond to the particle paths to detectors, the edge weights ($\in \mathbb{C}\setminus 0$) to the probability amplitudes, the edge colors ($\in \mathbb{Z}_2$) to the internal states of the particles along the paths. 

	
	The postselection of no-bunching states at the detectors correspond to the collection of $N$ edges that connect completely different vertices in the $G_{bb}$-representation. Such edge sets are called \emph{perfect matchings} (PMs) in graph theory (Appendix~\ref{glossary}). 
	The case of \eqref{N=2_bipartite} carries two perfect matchings, 
	\begin{align}\label{N=2_PMs}
		\begin{tikzpicture}[baseline={([yshift=-.5ex]current bounding box.center)}]
			\node[vertex] (1) at (0,2.5) {1};
			\node[vertex] (2) at (0,1) {2};  
			\node[circle,draw,minimum size=0.8cm] (X) at (3,2.5) {$X_1$};
			\node[circle,draw,minimum size=0.8cm] (Y) at (3,1) {$X_2$};	
			\path[line width = 0.8pt,color=blue] (1) edge   node[near start] {$\a_1$ } (X);
			\path[line width = 0.8pt,color=red] (2) edge  node[near start] {$\b_2$ } (Y);
			\draw (3.8,1.8) node[right] {+} ;
			\node[vertex] (1) at (4.5,2.5) {1};
			\node[vertex] (2) at (4.5,1) {2};  
			\node[circle,draw,minimum size=0.8cm] (X) at (7.5,2.5) {$X_1$};
			\node[circle,draw,minimum size=0.8cm] (Y) at (7.5,1) {$X_2$};		
			\path[line width = 0.8pt,color=blue] (1) edge   node[near start] {$\b_1$ } (Y);
			\path[line width = 0.8pt,color=red] (2) edge  node[near start] {$\a_2$ } (X);
		\end{tikzpicture},
	\end{align} which correspond to the two terms of Eq.~\eqref{N=2_fin}.

	\begin{widetext}
		\begin{center}
			\begin{table}[t]
				\begin{tabular}{|l|l|}
					\hline
					\textbf{Linear quantum network (LQN)}                & \textbf{Balanced bigraph $G_{bb}=(U\cup V, E)$}             \\ \hline \hline 
					$N$ identical particles & $N$ identical vertices  $\in U$ \\ \hline
					$N$ detectors & $N$ non-identical vertices  $\in V$   \\ \hline
					Possible paths of particles to arrive at detectors &  Edges $\in E$   \\ \hline
					Probability amplitudes & Edge weights ($\in\mathbb{C}\setminus 0$)\\ \hline 
					Internal states ($\uparrow$ and $\downarrow$) & Edge colors (B and R) \\ \hline 
					No-bunching states & Perfect matchings (PMs) \\ \hline
				\end{tabular}
				\caption{The correspondence relations between the elements of LQN and $G_{bb}$}
				\label{dictionary}
			\end{table}
		\end{center}
	\end{widetext}
	
	\twocolumngrid

	The correspondence relations between LQN and $G_{bb}$ are summarized in Table~\ref{dictionary}.
	
	\subsection{Computing the no-bunching states in $G_{bb}$-representation}\label{computation_protocol}
	
	By generalizing the procedure from Eq.~\eqref{N=2_bipartite} to Eq.~\eqref{N=2_PMs}, we can construct a systematic computation protocol for directly extracting the no-bunching states of a given LQN in the $G_{bb}$-representation:\\
	$ $\\
	Computation protocol
	\begin{enumerate}
		\item
		Using the mapping relation~\eqref{G_bb}, write down the adjacency matrices $A_{bb}$ and $C_{bb}$ of the $G_{bb}$-representation.
		\item
		Draw the corresponding $G_{bb}$.
		\item  
		Find all the PMs in the $G_{bb}$.
		\item
		Write down the no-bunching states corresponding to the PMs. For a no-bunching state, the internal state of the $j$th detector is determined by the color of the edge attached to $X_j$ in the PM.  The amplitude for the state is obtained by multiplying all the weights of the edges in the PM.
	\end{enumerate}
	Note that Step 1 can be skipped when the pictorial relation between the LQN and $G_{bb}$ is clear.
	
	By obtaining the relevant no-bunching terms directly from PMs in  the $G_{bb}$, the computation of the final postselected states become much simpler with the protocol.
	
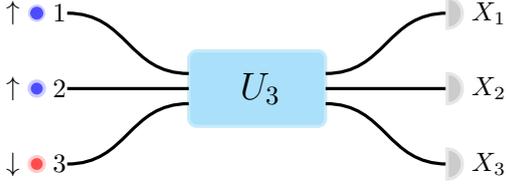
\begin{figure}
		\centering
		\begin{tikzpicture}
			\filldraw[rounded corners=3, very thick, color= cyan!20!white, fill=cyan!30!white] (.6,1) rectangle (2.4,2) ;
			\draw[fill=blue!70!white, thick, draw=blue!20!white ] (-1.4,2.5) circle (0.1);
			\draw[fill=blue!70!white, very thick, draw=blue!20!white ] (-1.4,1.5) circle (0.1) ;
			\draw[fill=red!70!white, very thick, draw=red!20!white ] (-1.4,0.5) circle (0.1) ;					
			\draw (1.15, 1.5)  node[right] {\Large $U_3$} ;
			\draw[very thick] (-1,0.5) .. controls (-.2,0.5) and (-.2,1.3) ..  (0.6,1.3); 
			\draw[very thick] (-1,1.5) -- (0.6,1.5);
			\draw[very thick] (-1,2.5) .. controls (-.2,2.5) and (-.2,1.7) .. (0.6,1.7);  
			\draw[very thick] (2.4,1.3) .. controls (3.2,1.3) and (3.2,.5) .. (4,0.5); 
			\draw[very thick] (2.4,1.5) -- (4,1.5);
			\draw[very thick] (2.4,1.7) .. controls (3.2,1.7) and (3.2,2.5) .. (4,2.5);       
			\draw (-1.5,0.5) node[left] {$\downarrow$}; 
			\draw (-1.5,1.5) node[left] {$\uparrow$}; 
			\draw (-1.5,2.5) node[left] {$\uparrow$}; 
			\draw (4.2,0.5) node[right] {$X_3$}; 
			\draw (4.2,1.5) node[right] {$X_2$}; 
			\draw (4.2,2.5) node[right] {$X_1$};
			\draw (-1.1,0.5) node {3}; 
			\draw (-1.1,1.5) node {2}; 
			\draw (-1.1,2.5) node {1}; 	
			\draw[fill=gray!40!white, very thick, draw=gray!20!white] (4,1.3) -- (4,1.7) arc(90:-90:.2) --cycle;                                  
			\draw[fill=gray!40!white, very thick, draw=gray!20!white] (4,2.3) -- (4,2.7) arc(90:-90:.2) --cycle;						   
			\draw[fill=gray!40!white, very thick, draw=gray!20!white] (4,.3) -- (4,.7) arc(90:-90:.2) --cycle;	      		    	
		\end{tikzpicture}
		\caption{An $N=3$ LQN with a tritter. Two bosons with $\uparrow$ are  injected to input modes 1 and 2, and a boson with $\downarrow$ is injected to 3. The total transformation matrix $U_3$ is unitary as Eq.~\eqref{U3}, and we postselect no-bunching states (one particle per output detector $X_j$, $j=1,2,3$).}
		\label{tritter}
	\end{figure}

	As an example, we compute a bosonic $N=3$ LQN whose transformation is given by a tritter (Figure~\ref{tritter}), which performs a balanced unitary transformation of three boson states with no internal state rotation. With two bosons initially in $\uparrow$ and one boson in $\downarrow$, the total state transformation relation with a tritter is written as
	\begin{align}\label{N=3_tritter}
		\begin{split}
			&\ha_{1\uparrow}\ha_{2\uparrow}\ha_{3\downarrow}|vac\> \\
			&\qquad \to  (u_{11}\ha_{1\uparrow}+u_{12}\ha_{2\uparrow}+u_{13}\ha_{3\uparrow}) \\
			&\qquad \times (u_{21}\ha_{1\uparrow}+u_{22}\ha_{2\uparrow}+u_{23}\ha_{3\uparrow}) \\
			&\qquad \times (u_{31}\ha_{1\downarrow}+u_{32}\ha_{2\downarrow}+u_{33}\ha_{3\downarrow})|vac\>,
		\end{split}
	\end{align}
	with the unitary transformation matrix
	\begin{align}\label{U3}
		U_3=
		\begin{pmatrix}
			u_{11}& u_{12} & u_{13} \\
			u_{21}& u_{22} & u_{23}\\
			u_{31}&u_{32}&u_{33}
		\end{pmatrix}=
		\frac{1}{\sqrt{3}}
		\begin{pmatrix}
			1& \w & \w^2 \\
			\w& 1 & \w^2\\
			1&1&1
		\end{pmatrix},
	\end{align} where $\w=e^{i\frac{2\pi}{3}}$.
	
	The expansion of Eq.~\eqref{N=3_tritter} contains $3^3=27$ terms. However, we can directly enumerate all the relevant no-bunching terms with our computation protocol as follows:

	$ $ \\	
	Steps 1 and 2. We can directly draw the corresponding $G_{bb}$ of Fig.~\ref{tritter} as
	\[\begin{tikzpicture}
		\node[vertex] (1) at (0,4) {$1 $} ;
		\node[vertex] (2) at (0,2) {$2 $};
		\node[vertex] (3) at (0,0) {$3 $};
		\node[circle,draw] (X) at (4,4) {$X_1$};
		\node[circle,draw] (Y) at (4,2) {$X_2$};
		\node[circle,draw] (Z) at (4,0) {$X_3$};	
		
		\path[line width = 0.8pt,color=blue] (1) edge   node[near start] {$u_{11}$ } (X);
		\path[line width = 0.8pt,color=blue] (1) edge   node[near start] {$u_{12}$ } (Y);
		\path[line width = 0.8pt,color=blue] (1) edge   node[near start] {$u_{13}$ } (Z);
		\path[line width = 0.8pt,color=blue] (2) edge   node[near start] {$u_{21}$ } (X);
		\path[line width = 0.8pt,color=blue] (2) edge   node[near start] {$u_{22}$ } (Y);
		\path[line width = 0.8pt,color=blue] (2) edge   node[near start] {$u_{23}$ } (Z);	
		\path[line width = 0.8pt,color=red] (3) edge   node[near start] {$u_{31}$ } (X);
		\path[line width = 0.8pt,color=red] (3) edge   node[near start] {$u_{32}$ } (Y);
		\path[line width = 0.8pt,color=red] (3) edge   node[near start] {$u_{33}$ } (Z);
		
	\end{tikzpicture}\]
	$ $ \\ Step 3. The above $G_b$ has six PMs, which are
	\[\begin{tikzpicture}
		\node[vertex] (1) at (0,2) {$1 $} ;
		\node[vertex] (2) at (0,1) {$2 $};
		\node[vertex] (3) at (0,0) {$3 $};
		\node[circle,draw] (X) at (2.5,2) {$X_1$};
		\node[circle,draw] (Y) at (2.5,1) {$X_2$};
		\node[circle,draw] (Z) at (2.5,0) {$X_3$};	
		\path[line width = 0.8pt,color=blue] (1) edge node[near start] {$u_{11}$} (X);
		\path[line width = 0.8pt,color=blue] (2) edge node[near start] {$u_{22}$} (Y);
		\path[line width = 0.8pt,color=red] (3) edge node[near start] {$u_{33}$} (Z);
		
		\node[vertex] (11) at (0,-1.5) {$1 $} ;
		\node[vertex] (22) at (0,-2.5) {$2 $};
		\node[vertex] (33) at (0,-3.5) {$3 $};
		\node[circle,draw] (XX) at (2.5,-1.5) {$X_1$};
		\node[circle,draw] (YY) at (2.5,-2.5) {$X_2$};
		\node[circle,draw] (ZZ) at (2.5,-3.5) {$X_3$};	
		\path[line width = 0.8pt,color=blue] (11) edge node[near start] {$u_{13}$} (ZZ);
		\path[line width = 0.8pt,color=blue] (22) edge node[near start] {$u_{21}$} (XX);
		\path[line width = 0.8pt,color=red] (33) edge node[near start] {$u_{32}$} (YY);
		
		\node[vertex] (11) at (0,-5) {$1$} ;
		\node[vertex] (22) at (0,-6) {$2 $};
		\node[vertex] (33) at (0,-7) {$3 $};
		\node[circle,draw] (XX) at (2.5,-5) {$X_1$};
		\node[circle,draw] (YY) at (2.5,-6) {$X_2$};
		\node[circle,draw] (ZZ) at (2.5,-7) {$X_3$};	
		\path[line width = 0.8pt,color=blue] (11) edge node[near start] {$u_{12}$} (YY);
		\path[line width = 0.8pt,color=blue] (22) edge node[near start] {$u_{23}$} (ZZ);
		\path[line width = 0.8pt,color=red] (33) edge node[near start] {$u_{31}$} (XX);

		\node[vertex] (1) at (4,2) {$1 $} ;
		\node[vertex] (2) at (4,1) {$2 $};
		\node[vertex] (3) at (4,0) {$3 $};
		\node[circle,draw] (X) at (6.5,2) {$X_1$};
		\node[circle,draw] (Y) at (6.5,1) {$X_2$};
		\node[circle,draw] (Z) at (6.5,0) {$X_3$};	
		\path[line width = 0.8pt,color=blue] (1) edge node[near start] {$u_{12}$} (Y);
		\path[line width = 0.8pt,color=blue] (2) edge node[near start] {$u_{21}$} (X);
		\path[line width = 0.8pt,color=red] (3) edge node[near start] {$u_{33}$} (Z);
		
		\node[vertex] (11) at (4,-1.5) {$1 $} ;
		\node[vertex] (22) at (4,-2.5) {$2 $};
		\node[vertex] (33) at (4,-3.5) {$3 $};
		\node[circle,draw] (XX) at (6.5,-1.5) {$X_1$};
		\node[circle,draw] (YY) at (6.5,-2.5) {$X_2$};
		\node[circle,draw] (ZZ) at (6.5,-3.5) {$X_3$};	
		\path[line width = 0.8pt,color=blue] (11) edge node[near start] {$u_{11}$} (XX);
		\path[line width = 0.8pt,color=blue] (22) edge node[near start] {$u_{23}$} (ZZ);
		\path[line width = 0.8pt,color=red] (33) edge node[near start] {$u_{32}$} (YY);
		
		\node[vertex] (11) at (4,-5) {$1$} ;
		\node[vertex] (22) at (4,-6) {$2 $};
		\node[vertex] (33) at (4,-7) {$3 $};
		\node[circle,draw] (XX) at (6.5,-5) {$X_1$};
		\node[circle,draw] (YY) at (6.5,-6) {$X_2$};
		\node[circle,draw] (ZZ) at (6.5,-7) {$X_3$};	
		\path[line width = 0.8pt,color=blue] (11) edge node[near start] {$u_{13}$} (ZZ);
		\path[line width = 0.8pt,color=blue] (22) edge node[near start] {$u_{22}$} (YY);
		\path[line width = 0.8pt,color=red] (33) edge node[near start] {$u_{31}$} (XX);	 	
		
	\end{tikzpicture}\]
	$ $\\ Step 4. By the exchange symmetry, the two PMs in the same line gives the same no-bunching terms. Therefore, the unnormalized final state at the level of detectors is given by
	\begin{align}
		|\P_{fin}\> =& 
		(u_{11}u_{22}+u_{12}u_{21})u_{33}|\uparrow_1\uparrow_2\downarrow_3\>\nn \\
		& + (u_{13}u_{21}+u_{11}u_{23})u_{32}|\uparrow_1\downarrow_2\uparrow_3\>\nn \\
		& +(u_{12}u_{23}+u_{13}u_{22})u_{31}|\downarrow_1\uparrow_2\uparrow_3\> \nn \\
		\sim& |\uparrow_1\uparrow_2\downarrow_3\> + \w|\uparrow_1\downarrow_2\uparrow_3\> +\w^2|\downarrow_1\uparrow_2\uparrow_3\>. 
	\end{align} 
	

	
	
	Finally, we have obtained \emph{a tripartite W state by extracting the relevant six terms}, not dealing with the whole $3^3=27$ terms that the LQN contains. This type of W state generation is realized with a linear optical system in Ref.~\cite{lee2021entangling}, which also studies the role of indistinguishability in generating genuinely entangled states.

	From now on, we suppose the particles are bosons for simplicity. A fermionic extension for the same LQN is easily achieved by applying the particle exchange $anti$-symmetry instead of the symmetry. 
	

	\subsection{Finding PMs in the $G_d$-representation}\label{G_d}	
	
	When our computation protocol is applied to an arbitrary $N$ particle LQN, the most non-trivial part is to enumerate all the PMs in the given $G_{bb}$. Although it was not very complicated for $N=2$ and $N=3$ examples we have discussed, we need a well-established protocol to find PMs in an LQN as the particle and edge numbers increase. For example, see the following $N=5$ bipartite graph
	\begin{align}\label{N=5}
		\begin{tikzpicture}[baseline={([yshift=-.5ex]current bounding box.center)}]
			\node[vertex] (1) at (6,2) {1} ;
			\node[vertex] (2) at (6,1) {2};
			\node[vertex] (3) at (6,0) {3};
			\node[vertex] (4) at (6,-1) {4}; 
			\node[vertex] (5) at (6,-2) {5};  	 
			\node[circle,draw,minimum size=0.8cm] (X) at (9,2) {$X_1$};
			\node[circle,draw,minimum size=0.8cm] (Y) at (9,1) {$X_2$};
			\node[circle,draw,minimum size=0.8cm] (Z) at (9,0) {$X_3$};	
			\node[circle,draw,minimum size=0.8cm] (W) at (9,-1) {$X_4$};
			\node[circle,draw,minimum size=0.8cm] (V) at (9,-2) {$X_5$};	    		    
			\path[line width = 0.8pt,color=red] (1) edge  (X);	
			\path[line width = 0.8pt,color=red] (1) edge  (W);
			\path[line width = 0.8pt,color=red] (2) edge  (X);
			\path[line width = 0.8pt,color=red] (2) edge  (Y);
			\path[line width = 0.8pt,color=red] (2) edge  (Z);
			\path[line width = 0.8pt,color=red] (2) edge  (W);
			\path[line width = 0.8pt,color=red] (2) edge    (V);				
			\path[line width = 0.8pt,color=blue] (3) edge  (X);
			\path[line width = 0.8pt,color=blue] (3) edge  (Z);	
			\path[line width = 0.8pt,color=blue] (4) edge  (W);
			\path[line width = 0.8pt,color=blue] (4) edge  (Z);
			\path[line width = 0.8pt,color=blue] (4) edge  (X);	
			\path[line width = 0.8pt,color=blue] (5) edge  (Y);
			\path[line width = 0.8pt,color=blue] (5) edge  (V);	 	
		\end{tikzpicture} \quad 
	\end{align} (whenever the edge weights are omitted in graphs, we presume they are set to be arbitrary $T_{aj}$ for $(a,X_j)$). It can be implemented in the optical setup with the second particle sent through $5\times 5$ multiport, the fourth through a $3\times 3$ multiport, and the others through beam splitters. 
	
	Here, we explain a systematic method to find PMs by mapping the $G_{bb}$ to a $G_{d}$ and examining the topology of the $G_d$. It was shown in Refs.~\cite{fukuda1994finding,tassa2012finding} that for a $G_{bb}$ with non-zero PMs, the existence of a PM is equivalent to that of an elementary cycle (a circular path that do not pass the same vertex twice, see Appendix~\ref{glossary}) in the $G_{d}$.
	We can use the relation to find all the no-bunching states of an LQN.
	
	We can draw a $G_d$ directly from a $G_b= (U\cup V, E)$ by mapping two vertices $i$ ($\in U$) and $X_i$ ($\in V$) to one vertex $w_i$ ($\in W$), and a undirected edge $(i, X_j)$ ($\in E $) to a directed edge $(w_i\to w_j)$ ($\in F$).
	For the case of \eqref{N=5}, the $G_d$ is  given by
	\begin{align}\label{N=5_Gd}
		\begin{tikzpicture}[baseline={([yshift=-.5ex]current bounding box.center)}]
			\node[circle,draw] (a) at (6,2.7) {$w_1$};
			\node[circle,draw] (b) at (7.5,1.5) {$w_2$};		
			\node[circle,draw] (c) at (7,0) {$w_3$};
			\node[circle,draw] (d) at (5,0) {$w_4$};
			\node[circle,draw] (e) at (4.5,1.5) {$w_5$};		
			\path[every loop/.style={min distance=10mm,in=45,out=135,looseness=5,color=red ,->},line width = 0.5pt] (a) edge [loop above]   (a);
			\path[every loop/.style={min distance=10mm,in=0,out=90,looseness=5,color=red},line width = 0.5pt,->] (b) edge [loop above] (b);
			\path[every loop/.style={min distance=10mm,in=270,out=0,looseness=5,color=blue ,->},line width = 0.5pt] (c) edge [loop above]  (c);
			\path[every loop/.style={min distance=10mm,in=180,out=270,looseness=5,color=blue ,->},line width = 0.5pt] (d) edge [loop above]  (d);
			\path[every loop/.style={min distance=10mm,in=90,out=180,looseness=5,color=blue ,->},line width = 0.5pt] (e) edge [loop above]  (e);				
			\path[line width = 0.8pt,color=red,->] (b) edge (a);
			\path[line width = 0.8pt,color=red,->] (a) edge[bend right=15] (d);	
			\path[line width = 0.8pt,color=red,->] (b) edge (c);
			\path[line width = 0.8pt,color=red,->] (b) edge (d);
			\path[line width = 0.8pt,color=red,->] (b) edge[bend right=15] (e);	
			\path[line width = 0.8pt,color=blue ,->] (c) edge (a);
			\path[line width = 0.8pt,color=blue ,->] (d) edge (c);
			\path[line width = 0.8pt,color=blue ,->] (d) edge[bend right=15] (a);
			\path[line width = 0.8pt,color=blue ,->] (e) edge[bend right=15] (b);	
		\end{tikzpicture}.
	\end{align} 
	
	We can consider a protocol to find PMs in a $G_{bb}$ by looking into elementary cycles in the $G_d$:
	
	$ $\\
	PM-finding protocol
	\begin{enumerate}
		\item 
		If there is more than one PM, we can label the vertices such that each vertex in $G_d$ has a loop without loss of generality. Then the $N$ loops constitute a PM, $\{(1,X_{1}),(2,X_{2}),\cdots, (N,X_{N})\}$.
		\item 
		If there is an elementary cycle 
		\[
		(w_{i_1}\to w_{i_2}\to \cdots \to w_{i_{k}} \to w_{i_1})
		\]
		in the $G_d$, then the edge set %
		\begin{align}\label{PMs}
			&\{\underbrace{(i_1,X_{i_2}),(i_2,X_{i_3}),\cdots, (i_{k-1},X_{i_{k}}),(i_{k},X_{i_1})}_{\textrm{edge exchange along the elementary cycle} },\nn \\
			&\qquad\qquad
			(i_{k+1},X_{i_{k+1}}),\cdots, (i_{N},X_{i_{N}})\}
			\nonumber
		\end{align} ($1\le i_q\le N$ for $1\le q \le N$, and $i_q= i_{p}$  if and only if $q=p$)	 is also a PM in $G_b$.
		\item
		Repeat Step 2 until we exhaust all the elementary cycles in the $G_{d}$. 
		Note that if there are elementary cycles that have no overlapping vertex with each other, a simultaneous edge exchange of such cycles also results in a PM. 
	\end{enumerate}

	$ $\\ Applying the above protocol to \eqref{N=5_Gd},
	we first see that the five loops attached to the vertices in \eqref{N=5_Gd} gives a PM, 
	\begin{align}
		\{(1,X_{1}),(2,X_{2}),(3,X_3),(4,X_4),(5,X_{5})\}.
	\end{align}
	There are three elementary cycles, 
	\begin{align}
		\begin{split}
			&(w_2 \to w_5\to w_2), 
			\\
			&(w_1\to w_4\to w_1), 
			\\
			&(w_1\to w_4 \to w_3 \to w_1),
		\end{split}
	\end{align}
	which give three PMs,
	\begin{align}
		&\{(2,X_{5}),(5,X_{2}),(1,X_1),(3,X_3),(4,X_{4})\}, \nn \\ 
		&\{(1,X_{4}),(4,X_{1}),(2,X_2),(3,X_3),(5,X_{5})\},\nn \\ 		&\{(1,X_{4}),(4,X_{3}),(3,X_1),(2,X_2),(5,X_{5})\}.	
	\end{align} 
	Note that $(w_2 \to w_5\to w_2)$ and $(w_1\to w_4\to w_1)$ do not share any vertex,   and neither do $(w_2 \to w_5\to w_2)$ and $(w_1\to w_4 \to w_3 \to w_1)$. Therefore,  simultaneous edge exchanges of them give two more PMs,
	\begin{align}
		&\{(1,X_{4}),(4,X_{1}),(2,X_5),(5,X_2),(3,X_{3})\},\nn \\	&\{(2,X_{5}),(5,X_{2}),(1,X_4),(4,X_3),(3,X_{1})\}.	
	\end{align}
	In total, \eqref{N=5_Gd} gives 6 PMs, which are drawn in the $G_{bb}$-representation as
	\begin{align}\label{N=5_PM}	
		\begin{tikzpicture}[baseline={([yshift=-.5ex]current bounding box.center)}]
			\node[vertex] (1) at (0,3.5) {$1 $} ;
			\node[vertex] (2) at (0,2.5) {$2 $};
			\node[vertex] (3) at (0,1.5) {$3 $};
			\node[vertex] (4) at (0,0.5) {$4 $};
			\node[vertex] (5) at (0,-0.5) {$5 $};		
			\node[circle,draw] (X) at (2.5,3.5) {$X_1$};
			\node[circle,draw] (Y) at (2.5,2.5) {$X_2$};
			\node[circle,draw] (Z) at (2.5,1.5) {$X_3$};
			\node[circle,draw] (V) at (2.5,0.5) {$X_4$};
			\node[circle,draw] (W) at (2.5,-0.5) {$X_5$};		
			\path[line width = 0.8pt,color=red] (1) edge  (X);
			\path[line width = 0.8pt,color=red ] (2) edge  (Y);
			\path[line width = 0.8pt,color=blue ] (3) edge (Z);
			\path[line width = 0.8pt,color=blue ] (4) edge (V);
			\path[line width = 0.8pt,color=blue ] (5) edge (W);	
			\node[vertex] (10) at (4,3.5) {$1 $} ;
			\node[vertex] (20) at (4,2.5) {$2 $};
			\node[vertex] (30) at (4,1.5) {$3 $};
			\node[vertex] (40) at (4,.5) {$4 $};
			\node[vertex] (50) at (4,-.5) {$5 $};		
			\node[circle,draw] (X0) at (6.5,3.5) {$X_1$};
			\node[circle,draw] (Y0) at (6.5,2.5) {$X_2$};
			\node[circle,draw] (Z0) at (6.5,1.5) {$X_3$};
			\node[circle,draw] (V0) at (6.5,.5) {$X_4$};
			\node[circle,draw] (W0) at (6.5,-.5) {$X_5$};		
			\path[line width = 0.8pt,color=red] (10) edge  (X0);
			\path[line width = 0.8pt,color=red ] (20) edge  (W0);
			\path[line width = 0.8pt,color=blue ] (30) edge (Z0);
			\path[line width = 0.8pt,color=blue ] (40) edge (V0);
			\path[line width = 0.8pt,color=blue ] (50) edge (Y0);
			\node[vertex] (11) at (0,-2) {$1 $} ;
			\node[vertex] (22) at (0,-3) {$2 $};
			\node[vertex] (33) at (0,-4) {$3 $};
			\node[vertex] (44) at (0,-5) {$4 $};
			\node[vertex] (55) at (0,-6) {$5 $};		
			\node[circle,draw] (XX) at (2.5,-2) {$X_1$};
			\node[circle,draw] (YY) at (2.5,-3) {$X_2$};
			\node[circle,draw] (ZZ) at (2.5,-4) {$X_3$};
			\node[circle,draw] (VV) at (2.5,-5) {$X_4$};
			\node[circle,draw] (WW) at (2.5,-6) {$X_5$};		
			\path[line width = 0.8pt,color=red ] (11) edge (VV);
			\path[line width = 0.8pt,color=red ] (22) edge  (YY);
			\path[line width = 0.8pt,color=blue ] (33) edge (ZZ);
			\path[line width = 0.8pt,color=blue ] (44) edge (XX);
			\path[line width = 0.8pt,color=blue ] (55) edge  (WW);	
			\node[vertex] (a) at (4,-2) {$1 $} ;
			\node[vertex] (b) at (4,-3) {$2 $};
			\node[vertex] (c) at (4,-4) {$3 $};
			\node[vertex] (d) at (4,-5) {$4 $};
			\node[vertex] (e) at (4,-6) {$5 $};		
			\node[circle,draw] (P) at (6.5,-2) {$X_1$};
			\node[circle,draw] (Q) at (6.5,-3) {$X_2$};
			\node[circle,draw] (R) at (6.5,-4) {$X_3$};
			\node[circle,draw] (S) at (6.5,-5) {$X_4$};
			\node[circle,draw] (T) at (6.5,-6) {$X_5$};		
			\path[line width = 0.8pt,color=red] (a) edge  (S);
			\path[line width = 0.8pt,color=red ] (b) edge  (Q);
			\path[line width = 0.8pt,color=blue ] (c) edge (P);
			\path[line width = 0.8pt,color=blue ] (d) edge (R);
			\path[line width = 0.8pt,color=blue ] (e) edge (T);			
			\node[vertex] (aa) at (0,-7.5) {$1 $} ;
			\node[vertex] (bb) at (0,-8.5) {$2 $};
			\node[vertex] (cc) at (0,-9.5) {$3 $};
			\node[vertex] (dd) at (0,-10.5) {$4 $};
			\node[vertex] (ee) at (0,-11.5) {$5 $};		
			\node[circle,draw] (P0) at (2.5,-7.5) {$X_1$};
			\node[circle,draw] (Q0) at (2.5,-8.5) {$X_2$};
			\node[circle,draw] (R0) at (2.5,-9.5) {$X_3$};
			\node[circle,draw] (S0) at (2.5,-10.5) {$X_4$};
			\node[circle,draw] (T0) at (2.5,-11.5) {$X_5$};		
			\path[line width = 0.8pt,color=red] (aa) edge  (S0);
			\path[line width = 0.8pt,color=red ] (bb) edge  (T0);
			\path[line width = 0.8pt,color=blue ] (cc) edge (R0);
			\path[line width = 0.8pt,color=blue ] (dd) edge (P0);
			\path[line width = 0.8pt,color=blue ] (ee) edge (Q0);				
			\node[vertex] (11) at (4,-7.5) {$1 $} ;
			\node[vertex] (22) at (4,-8.5) {$2 $};
			\node[vertex] (33) at (4,-9.5) {$3 $};
			\node[vertex] (44) at (4,-10.5) {$4 $};
			\node[vertex] (55) at (4,-11.5) {$5 $};		
			\node[circle,draw] (XX) at (6.5,-7.5) {$X_1$};
			\node[circle,draw] (YY) at (6.5,-8.5) {$X_2$};
			\node[circle,draw] (ZZ) at (6.5,-9.5) {$X_3$};
			\node[circle,draw] (VV) at (6.5,-10.5) {$X_4$};
			\node[circle,draw] (WW) at (6.5,-11.5) {$X_5$};		
			\path[line width = 0.8pt,color=red ] (11) edge (VV);
			\path[line width = 0.8pt,color=blue ] (55) edge  (YY);
			\path[line width = 0.8pt,color=blue ] (33) edge (XX);
			\path[line width = 0.8pt,color=blue ] (44) edge (ZZ);
			\path[line width = 0.8pt,color=red ] (22) edge  (WW);		
		\end{tikzpicture}.
	\end{align}

	Mapping the PMs into the no-bunching states in the LQN, the unnormalized form of the final no-bunching state is written as
	\begin{align}\label{N=5_fin}
		&|\P_{fin}\> \nn \\
		&= T_{11}T_{22}T_{33}T_{44}T_{55}|\downarrow_1\downarrow_2\uparrow_3\uparrow_4\uparrow_5\> \nn \\
		&\quad  +T_{11}T_{52}T_{33}T_{44}T_{25} |\downarrow_1\uparrow_2\uparrow_3\uparrow_4\downarrow_5\> \nn \\
		&\quad +(T_{41}T_{22}T_{33} + T_{31}T_{22}T_{43})T_{14}T_{55}|\uparrow_1\downarrow_2\uparrow_3\downarrow_4\uparrow_5\>  \nn \\
		&\quad  +(T_{41} T_{52}T_{33} +T_{31}T_{52}T_{43} ) T_{14} T_{25}|\uparrow_1\uparrow_2\uparrow_3\downarrow_4\downarrow_5\>. 
	\end{align} 
	
	
	As we have shown, we can schematically compute the postselected no-bunching state for any form of LQNs by combining the computation protocol in $G_{bb}$ and the above PM-finding protocol in $G_d$. 
	The above protocol considerably reduces the effort to PM-finding process~\cite{fukuda1994finding, uno1997algorithms,uno2001fast}, even if the protocol become inefficient to find all PMs as the number of vertices and edges increases~\cite{ausiello2012complexity}.
	
	\section{Perfect Matching diagram and the entanglement of linear quantum networks}\label{PM diagram}

	A crucial observation in the process of enumerating PMs is that not all the edges in $G_d$ are relevant for constituting the final no-bunching state. Considering the $N=5$ example \eqref{N=5} again, we can see that the edges $(2,X_1),(2,X_4)$, and $(2,X_5)$ are not included in any PM from the fact that they are not in any elementary cycle in \eqref{N=5_Gd}. 
	With this in consideration, we define a subgraph of the $G_d$ for a streamlined analysis of the entanglement generation by an LQN:
	\begin{definition}\label{D_c}
		(perfect matching diagram) For a given $G_d$, we define a ``perfect matching diagram'' (PM diagram, $\dc$) of the $G_d$ as a directed subgraph in which only the loops and the edges included the elementary cycles of the $G_d$ are retained. 
	\end{definition}	
	Then all the edges in a $\dc$ are included in the PMs of the LQN. Since a $\dc$ has no dummy particle path that are irrelevant to the no-bunching states, it captures the physical essence of the LQN. For example, the $\dc$ of \eqref{N=5_Gd} is drawn as
	\begin{align}\label{N=5_D_c}
		\begin{tikzpicture}[baseline={([yshift=-.5ex]current bounding box.center)}]
			\node[circle,draw] (a) at (6,2.7) {$w_1$};
			\node[circle,draw] (b) at (7.5,1.5) {$w_2$};		
			\node[circle,draw] (c) at (7,0) {$w_3$};
			\node[circle,draw] (d) at (5,0) {$w_4$};
			\node[circle,draw] (e) at (4.5,1.5) {$w_5$};		
			\path[every loop/.style={min distance=10mm,in=45,out=135,looseness=5,color=red ,->},line width = 0.5pt] (a) edge [loop above]   (a);
			\path[every loop/.style={min distance=10mm,in=0,out=90,looseness=5,color=red},line width = 0.5pt,->] (b) edge [loop above] (b);
			\path[every loop/.style={min distance=10mm,in=270,out=0,looseness=5,color=blue ,->},line width = 0.5pt] (c) edge [loop above]  (c);
			\path[every loop/.style={min distance=10mm,in=180,out=270,looseness=5,color=blue ,->},line width = 0.5pt] (d) edge [loop above]  (d);
			\path[every loop/.style={min distance=10mm,in=90,out=180,looseness=5,color=blue ,->},line width = 0.5pt] (e) edge [loop above]  (e);				
			\path[line width = 0.8pt,color=red,->] (a) edge[bend right=15] (d);	
			\path[line width = 0.8pt,color=red,->] (b) edge[bend right=15] (e);	
			\path[line width = 0.8pt,color=blue ,->] (c) edge (a);
			\path[line width = 0.8pt,color=blue ,->] (d) edge (c);
			\path[line width = 0.8pt,color=blue ,->] (d) edge[bend right=15] (a);
			\path[line width = 0.8pt,color=blue ,->] (e) edge[bend right=15] (b);
		\end{tikzpicture},
	\end{align} which gives the same perfect matchings as~\eqref{N=5_Gd}. 
	Finding the $\dc$ of an LQN is equivalent to finding maximally-matchable edges (edges in at least one of the PMs, see Appendix~\ref{glossary}) of a $G_{bb}$, which is efficiently achieved in  $\mathcal{O}(N+L)$ steps~\cite{tassa2012finding}. 
	
	Reversely, when we build an LQN to obtain some entangled no-bunching state, we should make its $G_d$ to be equal to $\dc$, i.e., all vertices in elementary cycles, to avoid dummy paths from the beginning.	
	
	In this section, we will present some rigorous relations between the structure of a $\dc$ and the entanglement of the total no-bunching state in the LQN. We show that for an LQN to generate a $N$-partite genuinely entangled state, the $\dc$ must be strongly connected and each vertex in it must have more than two incoming edges of different colors. We employ this relation to design LQNs fundamental genuinely entangled states. 

	\subsection{$\dc$ structure and the entanglement of LQNs}	\label{pmdiagram_entanglement}

	\begin{figure}
		\centering 	
		\begin{tikzpicture}
			\node (img) {\includegraphics[width=6cm]{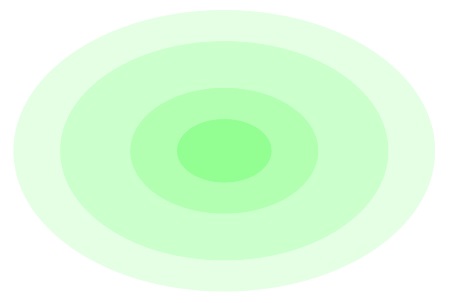}};
			\draw (0,0) node {\scriptsize $N$-sep.};
			\draw (0,-.56) node {\scriptsize $(N-1)$-sep.};	
			\draw (0,-1) node {$\vdots$};
			\draw (0,-1.66) node {\scriptsize $1$-sep.};	
		\end{tikzpicture}
		\caption{The inclusion relation of $k$-separable states. If a state is $k$-separable, then it is trivially $k-1$-separable. The states in the innermost circle is N-separable/entangled, i.e., \emph{fully separable}, and those in the outermost ring is $1$-separable/entangled, i.e., \emph{genuinely entangled.}
		}
		\label{ksep}
	\end{figure}
	
	Before scrutinizing the relation between the structure of $\dc$ and entanglement, we summarize the criteria for discriminating the hierarchy of pure state entanglement in multipartite systems. For a more thorough explanation including mixed states, see, e.g., Refs.~\cite{seevinck2008partial,szalay2012partial}.
	
	For a general $N$-partite system, the degree of entanglement for a pure state is widely categorized into three classes. First, a pure state $|\p\>$ ($\in \mH = \otimes_{j=1}^N \mH_i$) is \emph{fully separable} if it can be written as $|\p\> = |\p_1\>\otimes |\p_2\>\otimes \cdots \otimes |\p_N\>$ where $|\p_j\>\in \mathcal{H}_j$ for all $j=1,2,\cdots ,N$. Second, the state is \emph{genuinely (fully) entangled} if 
	$|\p\> \neq |\phi\>_{S}\otimes|\chi\>_{\bar{S}}$ for any bipartition $S|\bar{S}$ ($S\cup\bar{S}=\mH $  and $S\cap \bar{S}= \emptyset$). Third, the state is \emph{partially separable} when it is not fully separable but there exists at least one bipartition $S|\bar{S}$ of $\mH$ such that $|\p\> = |\phi\>_{S}\otimes|\chi\>_{\bar{S}}$.
	
	The partially separable states are divided into many classes according to the possible number of partitions. More specifically, a state is $k$-separable ($1\le k \le N$) if there exists a $k$-partition $\a_k =S_1|S_2|\cdots| S_k$ of $\mH$ ($S_1\cup S_2\cup \cdots \cup S_k=\mH $  and $S_a\cap S_b= \emptyset$ for any  $a$ and $b$ between 1 and $k$) such that $|\p\>$ is written as 
	$|\p\> = |\p_{S_1}\>\otimes |\p_{S_2}\>\otimes \cdots\otimes |\p_{S_k}\>$
	with $|\p_{S_a}\>\in S_{a}$ for $a=1,\cdots, k$. It is trivial to see that a $k$-separable state is always $(k-1)$-separable. If a state is $k$-separable but not $(k+1)$-separable, it is called \emph{$k$-separable/entangled}. 1-separable/entangled states are genuinely entangled, and $N$-separable/entangled states are fully separable. Fig.~\ref{ksep} shows the inclusion relation of the $k$-separability hierarchy.    

	We can show how the structure of a $\dc$ reveals the separability conditions of the no-bunching state in the LQN. First, one of the most obvious relations between the structure of $\dc$ and the entanglement of an LQN can be stated as the following lemma. 
	\begin{lemma}\label{separability}
		A subsystem $X_i$ of an LQN is separable from the other subsystems in the computational basis ($|\uparrow \> $ or  $|\downarrow \>$) if and only if all  incoming edges of the corresponding vertex $w_i$ in the $\dc$ are of the same color (a special case is when $w_i$ has only one incoming edge from a loop). 
	\end{lemma}
	\begin{proof}
		First, suppose that the incoming edges of $w_i$ have the same color. Then, according to the PM-finding protocol, the subsystem $X_i$ can have only one state ($|\uparrow\>$ or $|\downarrow\>$) in any PM
		Conversely, suppose that a final total no-bunching state is written as $|\P_{fin}\> = |s_{i}\>_{X_i} | \Phi\>_{\bar{X_i}}$ ($\bar{X}_i$ is the complementary subsystem of $X_i$, and $s_i=\uparrow$ or $\downarrow$). From the $\dc$ representation viewpoint, it means that all  the incoming vertices of $w_i$ have the same color irrespective of the number of vertices. 
	\end{proof}
	By the above lemma, we can directly see that the $N=5$ example~\eqref{N=5} has a separable subsystem $X_3$ in $|\uparrow\>$, for $w_3$ has only blue incoming edges in its corresponding $\dc$~\eqref{N=5_D_c}. Indeed, the explicit calculation in Eq.~\eqref{N=5_fin} shows that the subsystem $X_3$ is separable. 	
	
	Note that even if all the vertices of a $\dc$ are in multiple cycles with both colors of incoming edges, the corresponding LQN can have a separable subsystem in a superposed state of $|\uparrow\>$ and $|\downarrow\>$. For example, 		
	\begin{align}
		\begin{tikzpicture}[baseline={([yshift=-.5ex]current bounding box.center)}]
			\node[circle,draw] (1) at (3.6,1.7) {$w_1$};
			\node[circle,draw] (2) at (4.6,0) {$w_2$};
			\node[circle,draw] (3) at (2.6,0) {$w_3$};
			\path[every loop/.style={min distance=10mm,in=45,out=135,looseness=5},line width = 0.5pt,->,color=blue] (1) edge [loop above] (1);
			\path[every loop/.style={min distance=10mm,in=270,out=0,looseness=5},line width = 0.5pt,->,color=blue] (2) edge [loop above] (2);
			\path[every loop/.style={min distance=10mm,in=180,out=270,looseness=5},line width = 0.5pt, color=red,->] (3) edge [loop above]  (3);
			\path[line width = 0.8pt,->,color=blue] (1) edge[bend left] (2);
			\path[line width = 0.8pt,color=red,->] (2) edge[bend left] (1);
			\path[line width = 0.8pt,->,color=blue] (2) edge[bend left] (3);
			\path[line width = 0.8pt,color=red,->] (3) edge[bend left] (2);
			\path[line width = 0.8pt,->,color=blue] (1) edge (3);     
			\draw (1.6,1) node {$\Longrightarrow$} ;        			
			\node[vertex] (11) at (-2,2) {$1 $} ;
			\node[vertex] (22) at (-2,1) {$2 $};
			\node[vertex] (33) at (-2,0) {$3 $};
			\node[circle,draw] (XX) at (.5,2) {$X_1$};
			\node[circle,draw] (YY) at (.5,1) {$X_2$};
			\node[circle,draw] (ZZ) at (.5,0) {$X_3$};	
			\path[line width = 0.8pt,color=blue] (11) edge (XX);
			\path[line width = 0.8pt,color=blue] (22) edge (YY);
			\path[line width = 0.8pt,color=red] (33) edge (ZZ);
			\path[line width = 0.8pt,color=blue] (11) edge (YY);
			\path[line width = 0.8pt,color=red] (22) edge (XX);
			\path[line width = 0.8pt,color=red] (33) edge (YY);
			\path[line width = 0.8pt,color=blue] (22) edge (ZZ);
			\path[line width = 0.8pt,color=blue] (11) edge (ZZ);
		\end{tikzpicture} \nn 
	\end{align}
	generates a no-bunching state $T_{11}T_{22}T_{33}|\uparrow_1\uparrow_2\downarrow_3\>+T_{11}T_{23}T_{32}|\uparrow_1\downarrow_2\uparrow_3\> + T_{12}T_{21}T_{33}|\downarrow_1\uparrow_2\downarrow_3\> + T_{21}T_{32}T_{13}|\downarrow_1\downarrow_2\uparrow_3\>$.  
	If we set $T_{1i} = T_{2i} =1/\sqrt{3}$ and $T_{3i}=1/\sqrt{2}$ for $i=1,2,3$, the state becomes proportional to 
	$(|\uparrow_1\>+|\downarrow_1\>)(|\uparrow_2\downarrow_3\> +|\downarrow_2\uparrow_3\>)$.  
	
	Second, the connectivity of a $\dc$ also discloses a separability condition of the LQN:
	\begin{lemma}\label{disconnection}
		If a $\dc$ is disconnected as a $k$-partition of vertices $W_1|W_2|\cdots| W_k$, then the no-bunching state of the LQN is $k$-separable under $\a_k =S_1|S_2|\cdots| S_k$ where $S_i$ is mapped to $W_i$ ($i=1,\cdots,k$).  
	\end{lemma}
	\begin{proof}
		For a set of vertices $W_1$, we obtain PMs by taking edge exchanges only in $W_1$. Then the corresponding state is a tensor product of a $S_1$ state and a $\overline{S_1}$ state. By taking the same kind of edge exchanges for all $W_a$ ($\in \b_k$) and sum the corresponding states, we obtain a state  that is separable under $S_1|\overline{S}_1$. By repeating the same process for all $W_i$, we see that the no-bunching state is $k$-separable under $\a_k$.
	\end{proof}
	The $\dc$ \eqref{N=5_D_c} is also a good example for the application of the above lemma. Since \eqref{N=5_D_c} has two disconnected sets of vertices $\{w_1,w_3,w_4\}|\{w_2,w_5\}$, it must be biseparable under $(X_1,X_3,X_4)|(X_2,X_5)$. Indeed, Eq.~\eqref{N=5_fin} is rewritten in the tensor product form  as
	\begin{align}\label{N=5_arr}
		&|\P_{fin}\> \nn \\ 
		&= \Big[T_{11}T_{33}T_{44}|\downarrow_1 \uparrow_4\>+ T_{14}(T_{41}T_{33}+ T_{43}T_{31})|\uparrow_1 \downarrow_4\>\Big] \nn \\
		&\qquad \otimes |\uparrow_3\>\Big[T_{22}T_{55}|\downarrow_2\uparrow_5\> + T_{25}T_{52}|\uparrow_2 \downarrow_5\>\Big].
	\end{align}
	Therefore, applying Lemma \ref{separability} and \ref{disconnection}, we see that the $N=5$ LQN~\eqref{N=5} generates a $3$-separable state under a 3-partition $(X_1,X_4)|(X_3)|(X_2,X_5)$.
	
	Based on Lemma 1 and 2, we present the following theorem that provides necessary conditions for a $\dc$ with $N$ vertices to be genuinely $N$-partite entangled. 
	\begin{theorem}\label{genuine_ent}
		If an LQN generates a genuinely entangled no-bunching state, then i) each vertex in the $\dc$ must have more than two incoming edges of different colors, and ii) all the vertices in it are strongly connected to each other (strong connection of $w_i$ and $w_j$: we can move from $w_i$ toward $w_j$ and also from $w_j$ toward $w_i$). 
	\end{theorem}	
	\begin{proof}
		Since Lemma~\ref{separability} shows that the first condition is necessary, the remaining part of the proof is to show that the second condition is necessary. Using Lemma~\ref{disconnection}, all the vertices must be connected (i.e., have a path with each other) for an LQN to generate a genuinely entangled state. On the other hand, by Definition~\ref{D_c}, all the edges in a $\dc$ must be in elementary cycles. Combining Lemma~\ref{disconnection} and Definition~\ref{D_c}, all the elementary cycles in the $\dc$ must be connected, which results in the strong connection between any pair of vertices. Therefore, the second condition is necessary for the generation of genuine entanglement.    
	\end{proof}
	
	The above theorem is useful not only for checking the separability of a given LQN, but also for designing LQNs. Indeed, when one tries to design an LQN for a specific genuinely entangled state, the connectivity and color distribution of the $\dc$ provide solid guidelines.

	\subsection{Designing LQNs for genuinely entangled states in the $\dc$-representation}\label{applications}
	
	Now, we design $\dc$s that generate fundamental $N$-partite genuinely entangled states, i.e., states in GHZ and W classes, and the Dicke state.
	
	To achieve an arbitrary no-bunching state in an LQN that we expect, we can consider a ``trial and error'' method to construct a suitable $\dc$ that contains a given set of $L$ PMs, which can be summarized as follows:
	
	\begin{enumerate}
		\item 
		Among the $L$ PMs, choose one PM to be mapped to a $\dc$ of $N$ loops. 
		\item 
		Map another PM to a $\dc$ with an elementary cycle so that there is no discrepant loop or edge with the first $\dc$.
		\item
		Repeat until we obtain $L$ $\dc$s.
		\item 
		Pile up all the $\dc$s to obtain a $\dc$ that contains all the PMs. 
	\end{enumerate}

	It is guaranteed by construction that the final $\dc$ is mapped to an LQN that includes the $L$ PMs. However, such constructed $\dc$ may also have unwanted PMs that are not in the $L$ PMs. We call such a PM from the constructed $\dc$ a \emph{superfluous PM}. While it is not clear whether an universal strategy exists for avoiding the superfluous PMs, we can find optimal strategies for some specific states, especially those in GHZ and W classes. 
	
	In the following, we use the insight of Section~\ref{pmdiagram_entanglement} to construct LQNs for fundamental genuinely entangled states, i.e., those in the GHZ and W classes, and the Dicke state.
	
	
	\paragraph{GHZ class generation.} We start our discussion from the standard $N$-partite GHZ state,
	\begin{align}\label{ghz_standard}
		|GHZ\> = \frac{1}{\sqrt{2}}\big(|\uparrow_1\uparrow_2\cdots \uparrow_N\> + |\downarrow_1\downarrow_2\cdots \downarrow_N\>\big).
	\end{align}
	There are two PMs for the state,
	\begin{align}\label{NGHZ}
		\begin{tikzpicture}[baseline={([yshift=-.5ex]current bounding box.center)}]
			\node[vertex] (1) at (0,2) {1} ;
			\node[vertex] (2) at (0,1) {2};
			\node[vertex] (3) at (0,0) {3};
			\draw[dashed, thick] (0,-0.5) -- (0,-1.5); 
			\node[vertex] (4) at (0,-2) {N};    
			\node[circle,draw,minimum size=0.8cm] (X) at (2.5,2) {$X_1$};
			\node[circle,draw,minimum size=0.8cm] (Y) at (2.5,1) {$X_2$};
			\node[circle,draw,minimum size=0.8cm] (Z) at (2.5,0) {$X_3$};	
			\draw[dashed, thick] (2.5,-0.5) -- (2.5,-1.5); 	
			\node[circle,draw,minimum size=0.8cm] (W) at (2.5,-2) {$X_N$};	    
			\path[line width = 0.8pt,color=blue] (1) edge (X);
			\path[line width = 0.8pt,color=blue] (2) edge (Y);
			\path[line width = 0.8pt,color=blue] (3) edge (Z);
			\path[line width = 0.8pt,color=blue] (4) edge (W); 
			\draw[dashed, thick] (1.25,-1) -- (1.25,-1.4); 
			\node[vertex] (1) at (4,2) {1} ;
			\node[vertex] (2) at (4,1) {2};
			\node[vertex] (3) at (4,0) {3};
			\draw[dashed, thick] (4,-0.5) -- (4,-1.5); 
			\node[vertex] (4) at (4,-2) {N};    
			\node[circle,draw,minimum size=0.8cm] (X) at (6.5,2) {$X_1$};
			\node[circle,draw,minimum size=0.8cm] (Y) at (6.5,1) {$X_2$};
			\node[circle,draw,minimum size=0.8cm] (Z) at (6.5,0) {$X_3$};	
			\draw[dashed, thick] (6.5,-0.5) -- (6.5,-1.5); 	
			\node[circle,draw,minimum size=0.8cm] (W) at (6.5,-2) {$X_N$};	    
			\path[line width = 0.8pt,color=red] (1) edge (X);
			\path[line width = 0.8pt,color=red] (2) edge (Y);
			\path[line width = 0.8pt,color=red] (3) edge (Z);
			\path[line width = 0.8pt,color=red] (4) edge (W); 
			\draw[dashed, thick] (5.25,-1) -- (5.25,-1.4); 
		\end{tikzpicture}.
	\end{align}
	First, we choose the left one to be a $\dc$ of $N$ loops,
	\[\begin{tikzpicture}[baseline={([yshift=-.5ex]current bounding box.center)}]
		\node[circle,draw] (a) at (6,2.7) {$w_1$};
		\node[circle,draw] (b) at (7.5,1.5) {$w_2$};		
		\node[circle,draw] (c) at (7,0) {$w_3$};
		\node[circle,draw] (e) at (4.5,1.5) {$w_N$};		
		\path[every loop/.style={min distance=10mm,in=45,out=135,looseness=5,color=blue,->},line width = 0.5pt] (a) edge [loop above]   (a);
		\path[every loop/.style={min distance=10mm,in=0,out=90,looseness=5,color=blue},line width = 0.5pt,->] (b) edge [loop above] (b);
		\path[every loop/.style={min distance=10mm,in=270,out=0,looseness=5,color=blue ,->},line width = 0.5pt] (c) edge [loop above]  (c);
		\path[every loop/.style={min distance=10mm,in=90,out=180,looseness=5,color=blue ,->},line width = 0.5pt] (e) edge [loop above]  (e);	
		\draw[thick, dashed] (4.5,0.8) to[bend right=50] (6.3,-.3);				
	\end{tikzpicture}\] 
	Then, any $\dc$ with a elementary cycle that involves all the vertices can be the second $\dc$. One of such a $\dc$ is  
	\[\begin{tikzpicture}[baseline={([yshift=-.5ex]current bounding box.center)}]
		\node[circle,draw] (a) at (6,2.7) {$w_1$};
		\node[circle,draw] (b) at (7.5,1.5) {$w_2$};		
		\node[circle,draw] (c) at (7,0) {$w_3$};
		
		\node[circle,draw] (e) at (4.5,1.5) {$w_N$};		
		\path[line width = 0.8pt,color=red,->] (a) edge (b);
		\path[line width = 0.8pt,color=red,->] (b) edge (c);
		\path[line width = 0.8pt,color=red,->] (e) edge (a);						
		\draw[thick, dashed] (4.5,0.8) to[bend right=50] (6.3,-.3);				
	\end{tikzpicture}\] 
	By piling up the two $\dc$s, we obtain a $\dc$ that gives two PMs for the $N$-partite GHZ state,
	\begin{align}\label{GHZ_PM}
		&\begin{tikzpicture}[baseline={([yshift=-.5ex]current bounding box.center)}]
			\node[circle,draw] (a) at (6,2.7) {$w_1$};
			\node[circle,draw] (b) at (7.5,1.5) {$w_2$};		
			\node[circle,draw] (c) at (7,0) {$w_3$};
			\node[circle,draw] (e) at (4.5,1.5) {$w_N$};		
			\path[every loop/.style={min distance=10mm,in=45,out=135,looseness=5,color=blue,->},line width = 0.5pt] (a) edge [loop above]   (a);
			\path[every loop/.style={min distance=10mm,in=0,out=90,looseness=5,color=blue},line width = 0.5pt,->] (b) edge [loop above] (b);
			\path[every loop/.style={min distance=10mm,in=270,out=0,looseness=5,color=blue ,->},line width = 0.5pt] (c) edge [loop above]  (c);
			\path[every loop/.style={min distance=10mm,in=90,out=180,looseness=5,color=blue ,->},line width = 0.5pt] (e) edge [loop above]  (e);	
			\draw[thick, dashed] (4.5,0.8) to[bend right=50] (6.3,-.3);		
			\path[line width = 0.8pt,color=red,->] (a) edge (b);
			\path[line width = 0.8pt,color=red,->] (b) edge (c);
			\path[line width = 0.8pt,color=red,->] (e) edge (a);    					
		\end{tikzpicture}\nn \\
		&\quad \Longrightarrow \quad 
		\begin{tikzpicture}[baseline={([yshift=-.5ex]current bounding box.center)}]
			\node[vertex] (1) at (11,3) {1} ;
			\node[vertex] (2) at (11,2) {2};
			\node[vertex] (3) at (11,1) {3};
			\draw[dashed, thick] (11,.5) -- (11,-.5); 
			\node[vertex] (4) at (11,-1) {N};    
			\node[circle,draw,minimum size=0.8cm] (X) at (14,3) {$X_1$};
			\node[circle,draw,minimum size=0.8cm] (Y) at (14,2) {$X_2$};
			\node[circle,draw,minimum size=0.8cm] (Z) at (14,1) {$X_3$};	
			\draw[dashed, thick] (14,0.5) -- (14,-.5); 	
			\node[circle,draw,minimum size=0.8cm] (W) at (14,-1) {$X_N$};	    
			\path[line width = 0.8pt,color=blue] (1) edge (X);
			\path[line width = 0.8pt,color=blue] (2) edge (Y);
			\path[line width = 0.8pt,color=blue] (3) edge (Z);
			\path[line width = 0.8pt,color=blue] (4) edge (W); 
			\path[line width = 0.8pt,color=red] (1) edge (Y);
			\path[line width = 0.8pt,color=red] (2) edge (Z);
			\path[line width = 0.8pt,color=blue] (3) edge (Z);
			\path[line width = 0.8pt,color=red] (4) edge (X); 
			\draw[dashed, thick] (12.5,0) -- (12.5,.4); 
			\draw[line width = 0.8pt,color=red] (11.15,.95) -- (12.5,.5); 	
			\draw[line width = 0.8pt,color=red] (12.5,-.5) -- (13.55,-.85); 	
		\end{tikzpicture}.
	\end{align} 
	The above $\dc$ gives two PMs~\eqref{NGHZ} without a superfluous PM. The $N=3$ GHZ state generation as above  is experimentally given with a linear optical system in Ref~\cite{lee2021entangling}. 
	
	Extending the above analysis, we can conceive PM diagrams for the GHZ basis of $2^N$ states, i.e., states that can transform with local operations to the standard GHZ state~\eqref{ghz_standard}. Such PM diagrams has the same edge structure as \eqref{GHZ_PM} whose edge colors satisfy the necessary condition i) of Theorem~\ref{genuine_ent}.  
	By setting the edge colors as $c_j$ ($=B$ or $R$ for $j=1,\cdots, N$), and denoting $c_j \oplus 1$ as the different color from $c_j$, we can see that the  following $\dc$ constructs the GHZ class: 
	\begin{align}\label{GHZ_class}
		\begin{tikzpicture}[baseline={([yshift=-.5ex]current bounding box.center)}]
			\node[circle,draw] (a) at (0,3.5) {$w_1$};
			\node[circle,draw] (b) at (2,2) {$w_2$};		
			\node[circle,draw] (c) at (1.5,0) {$w_3$};
			\node[circle,draw] (e) at (-2,2) {$w_N$};		
			\path[every loop/.style={min distance=10mm,in=45,out=135,looseness=5,->},line width = 0.5pt] (a) edge [loop above] node[above] {$c_1$}   (a);
			\path[every loop/.style={min distance=10mm,in=0,out=90,looseness=5},line width = 0.5pt,->] (b) edge [loop above] node[above] {$c_2$} (b);
			\path[every loop/.style={min distance=10mm,in=270,out=0,looseness=5,->},line width = 0.5pt] (c) edge [loop above] node[below] {$c_3$} (c);
			\path[every loop/.style={min distance=10mm,in=90,out=180,looseness=5,->},line width = 0.5pt] (e) edge [loop above] node[above] {$c_N$} (e);	
			\draw[thick, dashed] (-1.9,1.2) to[bend right=55] (.7,-.2);		
			\path[line width = 0.8pt,->] (a) edge node[near start, right] {$c_2\oplus 1$} (b);
			\path[line width = 0.8pt,->] (b) edge  node[right] {$c_3\oplus 1$} (c);
			\path[line width = 0.8pt,->] (e) edge  node[near end, left] {$c_1 \oplus 1$} (a);    					
		\end{tikzpicture}. \nn 
	\end{align} The corresponding $G_{bb}$ becomes
	\begin{align} 
		\begin{tikzpicture}[baseline={([yshift=-.5ex]current bounding box.center)}]
			\node[vertex] (1) at (11,5) {1} ;
			\node[vertex] (2) at (11,3.5) {2};
			\node[vertex] (3) at (11,2) {3};
			\draw[dashed, thick] (11,1.5) -- (11,-.5); 
			\node[vertex] (4) at (11,-1) {N};    
			\node[circle,draw,minimum size=0.8cm] (X) at (14,5) {$X_1$};
			\node[circle,draw,minimum size=0.8cm] (Y) at (14,3.5) {$X_2$};
			\node[circle,draw,minimum size=0.8cm] (Z) at (14,2) {$X_3$};	
			\draw[dashed, thick] (14,1.3) -- (14,-.3); 	
			\node[circle,draw,minimum size=0.8cm] (W) at (14,-1) {$X_N$};	    
			\path[line width = 0.8pt] (1) edge node[near start] {$c_1$}  (X);
			\path[line width = 0.8pt] (2) edge node[near start] {$c_2$}  (Y);
			\path[line width = 0.8pt] (3) edge node[near start] {$c_3$}  (Z);
			\path[line width = 0.8pt] (4) edge node[near start] {$c_N$}  (W); 
			\path[line width = 0.8pt] (1) edge node {$c_2\oplus 1$}  (Y);
			\path[line width = 0.8pt] (2) edge node {$c_3\oplus 1$}  (Z);
			\path[line width = 0.8pt] (3) edge (Z);
			\path[line width = 0.8pt] (4) edge node[near start] {$c_1\oplus 1$}  (X); 
			\draw[dashed, thick] (12.5,1) -- (12.5,0); 
			\draw[line width = 0.8pt] (11.15,1.95) -- (12.5,1.27); 	
			\draw[line width = 0.8pt] (12.5,-.33) -- (13.55,-.85); 
			\draw (13,-0.5) node {$c_N\oplus 1$};	
		\end{tikzpicture}.
	\end{align} Indeed, the two PMs in \eqref{GHZ_class} correspond to 
	\begin{align}
		|c_1,c_2,\cdots, c_N\> + |c_1\oplus 1, c_1\oplus 1,\cdots , c_N\oplus 1\>. 
	\end{align} There are $2^N$ states of the above form, which construct the GHZ basis.  The $N=3$ GHZ basis example is explained 
	in Appendix~\ref{N=3_GHZW}.  
	
	\paragraph{W class generation.} The standard W state is a permutation symmetric state with one subsystem in $\downarrow$, i.e.,
	\begin{align}
		|W\> = \frac{1}{\sqrt{N}} \big(&|\downarrow_1\uparrow_2,\cdots,\uparrow_N\> + |\uparrow_1\downarrow_2\cdots \uparrow_N\>  \nn \\
		&+ \cdots +|\uparrow_1\uparrow_2\cdots \downarrow_N\>\big).
	\end{align}
	
	The expected $G_{bb}$ has $N$ PMs which contain one red edge per PM, i.e.,
	\begin{align}\label{NW}
		\begin{tikzpicture}[baseline={([yshift=-.5ex]current bounding box.center)}]
			\node[vertex] (1) at (0,2) {1} ;
			\node[vertex] (2) at (0,1) {2};
			\node[vertex] (3) at (0,0) {3};
			\draw[dashed, thick] (0,-0.5) -- (0,-1.5); 
			\node[vertex] (4) at (0,-2) {N};    
			\node[circle,draw,minimum size=0.8cm] (X) at (2.5,2) {$X_1$};
			\node[circle,draw,minimum size=0.8cm] (Y) at (2.5,1) {$X_2$};
			\node[circle,draw,minimum size=0.8cm] (Z) at (2.5,0) {$X_3$};	
			\draw[dashed, thick] (2.5,-0.5) -- (2.5,-1.5); 	
			\node[circle,draw,minimum size=0.8cm] (W) at (2.5,-2) {$X_N$};	    
			\path[line width = 0.8pt,color=red] (1) edge (X);
			\path[line width = 0.8pt,color=blue] (2) edge (Y);
			\path[line width = 0.8pt,color=blue] (3) edge (Z);
			\path[line width = 0.8pt,color=blue] (4) edge (W); 
			\draw[dashed, thick] (1.25,-1) -- (1.25,-1.4); 
		\end{tikzpicture},\quad 
		& \begin{tikzpicture}[baseline={([yshift=-.5ex]current bounding box.center)}]
			\node[vertex] (1) at (4,2) {1} ;
			\node[vertex] (2) at (4,1) {2};
			\node[vertex] (3) at (4,0) {3};
			\draw[dashed, thick] (4,-0.5) -- (4,-1.5); 
			\node[vertex] (4) at (4,-2) {N};    
			\node[circle,draw,minimum size=0.8cm] (X) at (6.5,2) {$X_1$};
			\node[circle,draw,minimum size=0.8cm] (Y) at (6.5,1) {$X_2$};
			\node[circle,draw,minimum size=0.8cm] (Z) at (6.5,0) {$X_3$};	
			\draw[dashed, thick] (6.5,-0.5) -- (6.5,-1.5); 	
			\node[circle,draw,minimum size=0.8cm] (W) at (6.5,-2) {$X_N$};	    
			\path[line width = 0.8pt,color=blue] (1) edge (X);
			\path[line width = 0.8pt,color=red] (2) edge (Y);
			\path[line width = 0.8pt,color=blue] (3) edge (Z);
			\path[line width = 0.8pt,color=blue] (4) edge (W); 
			\draw[dashed, thick] (5.25,-1) -- (5.25,-1.4); 
		\end{tikzpicture}, \nn \\ 
		\qquad \cdots\cdots, \qquad \quad  \quad 
		&\begin{tikzpicture}[baseline={([yshift=-.5ex]current bounding box.center)}]
			\node[vertex] (1) at (0,2) {1} ;
			\node[vertex] (2) at (0,1) {2};
			\node[vertex] (3) at (0,0) {3};
			\draw[dashed, thick] (0,-0.5) -- (0,-1.5); 
			\node[vertex] (4) at (0,-2) {N};    
			\node[circle,draw,minimum size=0.8cm] (X) at (2.5,2) {$X_1$};
			\node[circle,draw,minimum size=0.8cm] (Y) at (2.5,1) {$X_2$};
			\node[circle,draw,minimum size=0.8cm] (Z) at (2.5,0) {$X_3$};	
			\draw[dashed, thick] (2.5,-0.5) -- (2.5,-1.5); 	
			\node[circle,draw,minimum size=0.8cm] (W) at (2.5,-2) {$X_N$};	    
			\path[line width = 0.8pt,color=blue] (1) edge (X);
			\path[line width = 0.8pt,color=blue] (2) edge (Y);
			\path[line width = 0.8pt,color=blue] (3) edge (Z);
			\path[line width = 0.8pt,color=red] (4) edge (W); 
			\draw[dashed, thick] (1.25,-1) -- (1.25,-1.4); 
		\end{tikzpicture} .
	\end{align}
	
	There are several options to draw LQNs for the above set of PMs. By analyzing the $\dc$ of such LQNs, we can find a crucial property for any LQN that generates an $N$-partite W state with no superfluous term, which we state in the language of $\dc$ as the following theorem:   
	\begin{theorem}\label{wstate} Suppose that a $\dc$ is optimal for the $N$-partite W state, i.e., in the $\dc$, all the PMs for the $N$-partite W-state appear once with no superfluous PM.
		Then, the $\dc$ must have $N$ red edges that leave the same vertex, i.e, the $N$ red edges must distribute without loss of generality as 
		\begin{align}\label{redN}
			\begin{tikzpicture}[baseline={([yshift=-.5ex]current bounding box.center)}]
				\node[circle,draw] (a) at (6,2.7) {$w_1$};
				\node[circle,draw] (b) at (8,1) {$w_2$};		
				\node[circle,draw] (c) at (7,0) {$w_3$};
				\node[circle,draw] (e) at (4,1) {$w_N$};		
				\path[every loop/.style={min distance=10mm,in=45,out=135,looseness=5,color=red,->},line width = 0.5pt] (a) edge [loop above]   (a);
				\draw[thick, dashed] (4.3,0.3) to[bend right=30] (6.3,-.3);	
				\path[line width = 0.8pt,->,color=red] (a) edge  (b);
				\path[line width = 0.8pt,->,color=red] (a) edge  (c);
				\path[line width = 0.8pt,->,color=red] (a) edge  (e);
			\end{tikzpicture}.
		\end{align}	
	\end{theorem}
	\begin{proof}
		First, we can map the first PM of \eqref{NW} to $N$ loops,
		\begin{align}\label{wstate_loops}
			\begin{tikzpicture}[baseline={([yshift=-.5ex]current bounding box.center)}]
				\node[circle,draw] (a) at (6,2.7) {$w_1$};
				\node[circle,draw] (b) at (7.5,1.5) {$w_2$};		
				\node[circle,draw] (c) at (7,0) {$w_3$};
				\node[circle,draw] (e) at (4.5,1.5) {$w_N$};		
				\path[every loop/.style={min distance=10mm,in=45,out=135,looseness=5,color=red,->},line width = 0.5pt] (a) edge [loop above]   (a);
				\path[every loop/.style={min distance=10mm,in=0,out=90,looseness=5,color=blue},line width = 0.5pt,->] (b) edge [loop above] (b);
				\path[every loop/.style={min distance=10mm,in=270,out=0,looseness=5,color=blue ,->},line width = 0.5pt] (c) edge [loop above]  (c);
				\path[every loop/.style={min distance=10mm,in=90,out=180,looseness=5,color=blue ,->},line width = 0.5pt] (e) edge [loop above]  (e);	
				\draw[thick, dashed] (4.5,0.8) to[bend right=50] (6.3,-.3);		 	
			\end{tikzpicture}.
		\end{align}		
		Since the edge $(1,X_1)$ is red in only one PM, all the elementary cycles must pass $w_1$ to exchange the red loop of $w_1$ with other incoming blue edges. 
		In addition, the optimal $\dc$ for the W state has $(N-1)$-elementary cycles and each elementary cycle must have a different red edge from those in the other elementary cycles. 
		
		Using the above restrictions, we first show that if $(w_1\to w_j)$  exists for some $j$ ($2\le j \le N$), then it must be red, i.e., $(w_1\xrightarrow{R}w_j)$. Suppose  $(w_1 \xrightarrow{B} w_j)$. Then, since $w_j$ must have one red incoming edge, there exists an edge $(w_l\xrightarrow{R}w_j)$ where $l$ is neither $1$ nor $j$.
		On the other hand, $w_l$ also must have one red incoming edge. From the fact that $w_1$ is in all elementary cycles, we see that there exists at least one elementary cycle that includes the two red edges.  This contradicts with the restriction that each elementary cycle must have only one red edge. Therefore, $(w_1\to w_j)$ must be red.
		
		Now, we show that 
		$(w_1\to w_j)$ exists for all $j$ ($=2,\cdots,N$).  Suppose that an edge $(w_1\to w_k)$ ($2\le k \le N$) is absent. Then, considering there must be ($N-1$)-elementary cycles that pass $w_1$, an edge $(w_1 \to w_l)$ ($2\le l\le N$) must be included in more than one elementary cycle. Since no two elementary cycles can have the same red edge, $(w_1 \to w_k)$ must be blue, i.e., $(w_1 \xrightarrow{B} w_k)$, which is not possible as we have shown. Therefore, the $N$ red edges must distribute as \eqref{redN}. 
	\end{proof}
	According to the above theorem, we see that an LQN that carries a W state must have one identical particle that can be observed at all detectors as $\downarrow$, while all other particles are observed as $\uparrow$.
	
	Using  Theorems~\ref{genuine_ent} and \ref{wstate}, we suggest two ways to design LQNs for the W state.
	The first one consists only of $(w_1\xrightarrow{R} w_a\xrightarrow{B} w_1)$ where $a=2,\cdots, N$. By piling up such $(N-1)$-elementary cycles and \eqref{wstate_loops}, we obtain a $\dc$, 
	\begin{align}\label{wstate_1}
		\begin{tikzpicture}[baseline={([yshift=-.5ex]current bounding box.center)}]
			\node[circle,draw] (a) at (6,2.7) {$w_1$};
			\node[circle,draw] (b) at (8,1) {$w_2$};		
			\node[circle,draw] (c) at (7,0) {$w_3$};
			\node[circle,draw] (e) at (4,1) {$w_N$};		
			\path[every loop/.style={min distance=10mm,in=45,out=135,looseness=5,color=red,->},line width = 0.5pt] (a) edge [loop above]   (a);
			\path[every loop/.style={min distance=10mm,in=-45,out=45,looseness=5,color=blue},line width = 0.5pt,->] (b) edge [loop above] (b);
			\path[every loop/.style={min distance=10mm,in=270,out=0,looseness=5,color=blue ,->},line width = 0.5pt] (c) edge [loop above]  (c);
			\path[every loop/.style={min distance=10mm,in=135,out=-135,looseness=5,color=blue ,->},line width = 0.5pt] (e) edge [loop above]  (e);	
			\draw[thick, dashed] (4.3,0.3) to[bend right=30] (6.3,-.3);	
			\path[line width = 0.8pt,->,color=red] (a) edge [bend right=10] (b);
			\path[line width = 0.8pt,->,color=red] (a) edge [bend right=10] (c);
			\path[line width = 0.8pt,->,color=red] (a) edge [bend right=10] (e);
			\path[line width = 0.8pt,->,color=blue] (b) edge [bend right=10] (a);
			\path[line width = 0.8pt,->,color=blue] (c) edge [bend right=10] (a);
			\path[line width = 0.8pt,->,color=blue] (e) edge [bend right=10] (a);							
		\end{tikzpicture}.
	\end{align}
	We see that each elementary cycle includes one red edge, which gives a PM that composes the $N$-partite W state.
	
	Another $\dc$ can contain an elementary cycle of $N$ vertices $(w_1\xrightarrow{R}w_N\xrightarrow{B}\cdots \xrightarrow{B} w_2 \xrightarrow{B} w_1)$,
	\begin{align}
		\begin{tikzpicture}[baseline={([yshift=-.5ex]current bounding box.center)}]
			\node[circle,draw] (a) at (6,2.7) {$w_1$};
			\node[circle,draw] (b) at (8,1.5) {$w_2$};		
			\node[circle,draw] (c) at (7,0) {$w_3$};
			\node[circle,minimum size =0.9cm, draw] (d) at (5,0) { };
			\node[circle,draw] (e) at (4,1.5) {$w_N$};		
			\path[line width = 0.8pt,color=red,->] (a) edge[bend right=10]  (e);
			\path[line width = 0.8pt,color=blue,->] (e) edge[bend right=10] (d);
			\path[line width = 0.8pt,color=blue,->] (c) edge[bend right=10] (b);
			\path[line width = 0.8pt,color=blue,->] (b) edge[bend right=10] (a);
			\draw[thick, dashed] (5.6,-0.3) to[bend right=10] (6.5,-.3);		
			\draw (5,0) node {$w_{N-1}$};	
		\end{tikzpicture}.
	\end{align}
	By piling up \eqref{redN} with the above, we have
	\begin{align}\label{wstate_2}
		\begin{tikzpicture}[baseline={([yshift=-.5ex]current bounding box.center)}]
			\node[circle,draw] (a) at (6,2.7) {$w_1$};
			\node[circle,draw] (b) at (8,1.5) {$w_2$};		
			\node[circle,draw] (c) at (7,0) {$w_3$};
			\node[circle,minimum size =0.9cm, draw] (d) at (5,0) { };
			\node[circle,draw] (e) at (4,1.5) {$w_N$};	
			\path[every loop/.style={min distance=10mm,in=45,out=135,looseness=5,color=red ,->},line width = 0.5pt] (a) edge [loop above]   (a);
			\path[every loop/.style={min distance=10mm,in=-45,out=45,looseness=5,color=blue},line width = 0.5pt,->] (b) edge [loop above] (b);
			\path[every loop/.style={min distance=10mm,in=270,out=0,looseness=5,color=blue ,->},line width = 0.5pt] (c) edge [loop above]  (c);
			\path[every loop/.style={min distance=10mm,in=185,out=265,looseness=5,color=blue ,->},line width = 0.5pt] (d) edge [loop above]  (d);
			\path[every loop/.style={min distance=10mm,in=135,out=-135,looseness=5,color=blue ,->},line width = 0.5pt] (e) edge [loop above]  (e);				
			\path[line width = 0.8pt,color=red,->] (a) edge[bend right=10]  (e);
			\path[line width = 0.8pt,color=blue,->] (e) edge[bend right=10] (d);
			\path[line width = 0.8pt,color=blue,->] (c) edge[bend right=10] (b);
			\path[line width = 0.8pt,color=blue,->] (b) edge[bend right=10] (a);
			\path[line width = 0.8pt,color=red,->] (a) edge  (d);
			\path[line width = 0.8pt,color=red,->] (a) edge  (c);
			\path[line width = 0.8pt,color=red,->] (a) edge[bend right=10]  (b);
			\draw[thick, dashed] (5.6,-0.3) to[bend right=10] (6.5,-.3);		
			\draw (5,0) node {$w_{N-1}$};	
		\end{tikzpicture}.
	\end{align} 
	The ($N-1$)-elementary cycles are of the form
	$(w_1\xrightarrow{R}w_j\xrightarrow{B}w_{j-1}\xrightarrow{B}\cdots \xrightarrow{B}w_2\xrightarrow{B}w_1)$ where $j=2,\cdots, N$, which also gives PMs that construct the $N$-partite W state.  It is direct to see that \eqref{wstate_1} and \eqref{wstate_2} satisfy the neccessary conditions of Theorems~\ref{genuine_ent} and \ref{wstate}. 
	
	For example, when $N=5$, \eqref{wstate_1} and its corresponding $G_{bb}$ are given by
	\begin{align}
		&\begin{tikzpicture}[baseline={([yshift=-.5ex]current bounding box.center)}]
			\node[circle,draw] (a) at (6,3.3) {$w_1$};
			\node[circle,draw] (b) at (8.5,1.5) {$w_2$};		
			\node[circle,draw] (c) at (7,0.7) {$w_3$};
			\node[circle,draw] (d) at (5,0.7) {$w_4$};
			\node[circle,draw] (e) at (3.5,1.5) {$w_5$};		
			\path[every loop/.style={min distance=10mm,in=45,out=135,looseness=5,color=red ,->},line width = 0.5pt] (a) edge [loop above]   (a);
			\path[every loop/.style={min distance=10mm,in=0,out=90,looseness=5,color=blue},line width = 0.5pt,->] (b) edge [loop above] (b);
			\path[every loop/.style={min distance=10mm,in=270,out=0,looseness=5,color=blue ,->},line width = 0.5pt] (c) edge [loop above]  (c);
			\path[every loop/.style={min distance=10mm,in=180,out=270,looseness=5,color=blue ,->},line width = 0.5pt] (d) edge [loop above]  (d);
			\path[every loop/.style={min distance=10mm,in=90,out=180,looseness=5,color=blue ,->},line width = 0.5pt] (e) edge [loop above]  (e);				
			\path[line width = 0.8pt,color=red,->] (a) edge[bend right=15] (b);
			\path[line width = 0.8pt,color=blue,->] (b) edge[bend right=15] (a);	
			\path[line width = 0.8pt,color=red,->] (a) edge[bend right=15] (c);
			\path[line width = 0.8pt,color=blue,->] (c) edge[bend right=15] (a);
			\path[line width = 0.8pt,color=red,->] (a) edge[bend right=15] (d);
			\path[line width = 0.8pt,color=blue,->] (d) edge[bend right=15] (a);
			\path[line width = 0.8pt,color=red,->] (a) edge[bend right=15] (e);
			\path[line width = 0.8pt,color=blue,->] (e) edge[bend right=15] (a);
		\end{tikzpicture} \nn \\
		& \qquad \Longrightarrow \quad 
		\begin{tikzpicture}[baseline={([yshift=-.5ex]current bounding box.center)}]      	
			\node[vertex] (1) at (12,4) {$1$} ;
			\node[vertex] (2) at (12,3) {$2$};
			\node[vertex] (3) at (12,2) {$3$};
			\node[vertex] (4) at (12,1) {$4$}; 
			\node[vertex] (5) at (12,0) {$5$};  	 
			\node[circle,draw,minimum size=0.8cm] (X) at (15,4) {$X_1$};
			\node[circle,draw,minimum size=0.8cm] (Y) at (15,3) {$X_2$};
			\node[circle,draw,minimum size=0.8cm] (Z) at (15,2) {$X_3$};	
			\node[circle,draw,minimum size=0.8cm] (W) at (15,1) {$X_4$};
			\node[circle,draw,minimum size=0.8cm] (V) at (15,0) {$X_5$};	    		    
			\path[line width = 0.8pt,color=red] (1) edge (Y);
			\path[line width = 0.8pt,color=red] (1) edge (Z);
			\path[line width = 0.8pt,color=red] (1) edge (W);
			\path[line width = 0.8pt,color=red] (1) edge (V);
			\path[line width = 0.8pt,color=red] (1) edge (X);	
			\path[line width = 0.8pt,color=blue ] (2) edge (X);	
			\path[line width = 0.8pt,color=blue ] (2) edge (Y);			
			\path[line width = 0.8pt,color=blue ] (3) edge (X);
			\path[line width = 0.8pt,color=blue ] (3) edge (Z);	
			\path[line width = 0.8pt,color=blue ] (4) edge (X);
			\path[line width = 0.8pt,color=blue ] (4) edge (W);
			\path[line width = 0.8pt,color=blue ] (5) edge (X);
			\path[line width = 0.8pt,color=blue ] (5) edge (V);	 
		\end{tikzpicture} 
	\end{align}
	And \eqref{wstate_2} and its corresponding $G_{bb}$ for $N=5$ are given by
	\begin{align}
		&	\begin{tikzpicture}[baseline={([yshift=-.5ex]current bounding box.center)}]
			\node[circle,draw] (a) at (6,3.3) {$w_1$};
			\node[circle,draw] (b) at (8.5,1.5) {$w_2$};		
			\node[circle,draw] (c) at (7,0.7) {$w_3$};
			\node[circle,draw] (d) at (5,0.7) {$w_4$};
			\node[circle,draw] (e) at (3.5,1.5) {$w_5$};		
			\path[every loop/.style={min distance=10mm,in=45,out=135,looseness=5,color=red ,->},line width = 0.5pt] (a) edge [loop above]   (a);
			\path[every loop/.style={min distance=10mm,in=0,out=90,looseness=5,color=blue},line width = 0.5pt,->] (b) edge [loop above] (b);
			\path[every loop/.style={min distance=10mm,in=270,out=0,looseness=5,color=blue ,->},line width = 0.5pt] (c) edge [loop above]  (c);
			\path[every loop/.style={min distance=10mm,in=180,out=270,looseness=5,color=blue ,->},line width = 0.5pt] (d) edge [loop above]  (d);
			\path[every loop/.style={min distance=10mm,in=90,out=180,looseness=5,color=blue ,->},line width = 0.5pt] (e) edge [loop above]  (e);				
			\path[line width = 0.8pt,color=red,->] (a) edge[bend right=15] (b);
			\path[line width = 0.8pt,color=blue,->] (b) edge[bend right=15] (a);	
			\path[line width = 0.8pt,color=red,->] (a) edge (c);
			\path[line width = 0.8pt,color=blue,->] (c) edge (b);
			\path[line width = 0.8pt,color=red,->] (a) edge (d);
			\path[line width = 0.8pt,color=blue,->] (d) edge (c);
			\path[line width = 0.8pt,color=red,->] (a) edge (e);
			\path[line width = 0.8pt,color=blue,->] (e) edge (d);
		\end{tikzpicture} \nn \\
		&\qquad \Longrightarrow\quad  
		\begin{tikzpicture}[baseline={([yshift=-.5ex]current bounding box.center)}]
			\node[vertex] (1) at (12,4) {$1$} ;
			\node[vertex] (2) at (12,3) {$2$};
			\node[vertex] (3) at (12,2) {$3$};
			\node[vertex] (4) at (12,1) {$4$}; 
			\node[vertex] (5) at (12,0) {$5$};  	 
			\node[circle,draw,minimum size=0.8cm] (X) at (15,4) {$X_1$};
			\node[circle,draw,minimum size=0.8cm] (Y) at (15,3) {$X_2$};
			\node[circle,draw,minimum size=0.8cm] (Z) at (15,2) {$X_3$};	
			\node[circle,draw,minimum size=0.8cm] (W) at (15,1) {$X_4$};
			\node[circle,draw,minimum size=0.8cm] (V) at (15,0) {$X_5$};	    		    
			\path[line width = 0.8pt,color=red] (1) edge (Y);
			\path[line width = 0.8pt,color=red] (1) edge (Z);
			\path[line width = 0.8pt,color=red] (1) edge (W);
			\path[line width = 0.8pt,color=red] (1) edge (V);
			\path[line width = 0.8pt,color=red] (1) edge (X);	
			\path[line width = 0.8pt,color=blue ] (2) edge (X);	
			\path[line width = 0.8pt,color=blue ] (2) edge (Y);			
			\path[line width = 0.8pt,color=blue ] (3) edge (Y);
			\path[line width = 0.8pt,color=blue ] (3) edge (Z);	
			\path[line width = 0.8pt,color=blue ] (4) edge (Z);
			\path[line width = 0.8pt,color=blue ] (4) edge (W);
			\path[line width = 0.8pt,color=blue ] (5) edge (W);
			\path[line width = 0.8pt,color=blue ] (5) edge (V);	 
		\end{tikzpicture}. 
	\end{align}
	
	It is worth mentioning other forms of LQNs that can generate the W state. 
	The $\dc$ of the tritter LQN (Eq.~\eqref{U3}) generates a W state by including the same PM twice. We can also generate a W state using an ancilla as suggested in Refs.~\cite{bellomo2017n,kim2020efficient, blasiak2019entangling}. 
	
	Similar to the GHZ case, we can consider PM diagrams that build the W basis of $2^N$ states. One group of the PM diagrams has the same edge structure as \eqref{wstate_1}, and the other has the same structure as \eqref{wstate_2}. In each group, the edge colors satisfy the necessary condition i) of Theorem~\ref{genuine_ent}. 
	
	The first group is of the form
	\begin{align}\label{wbasis_1}
		\begin{tikzpicture}[baseline={([yshift=-.5ex]current bounding box.center)}]
			\node[circle,draw] (a) at (6,2.7) {$w_1$};
			\node[circle,draw] (b) at (8.5,1.7) {$w_2$};		
			\node[circle,draw] (c) at (7.3,0) {$w_3$};
			\node[circle,draw] (e) at (3.5,1.7) {$w_N$};		
			\path[every loop/.style={min distance=10mm,in=45,out=135,looseness=5,->},line width = 0.5pt] (a) edge [loop above] node[above] {$c_1\oplus 1$}  (a);
			\path[every loop/.style={min distance=10mm,in=-45,out=45,looseness=5},line width = 0.5pt,->] (b) edge [loop above] node[right] {$c_2$} (b);
			\path[every loop/.style={min distance=10mm,in=270,out=0,looseness=5,->},line width = 0.5pt] (c) edge [loop above] node[below] {$c_3$} (c);
			\path[every loop/.style={min distance=10mm,in=135,out=-135,looseness=5,->},line width = 0.5pt] (e) edge [loop above] node[left] {$c_N$} (e);	
			\draw[thick, dashed] (3.7,0.7) to[bend right=30] (6.3,-.3);	
			\path[line width = 0.8pt,->] (a) edge [bend right=10] node {$c_2\oplus 1$} (b);
			\path[line width = 0.8pt,->] (a) edge [bend right=10] node[near end, left] {$c_3\oplus 1$} (c);
			\path[line width = 0.8pt,->] (a) edge [bend right=10] node[above] {$c_N\oplus 1$} (e);
			\path[line width = 0.8pt,->] (b) edge [bend right=10] node[above] {$c_1$} (a);
			\path[line width = 0.8pt,->] (c) edge [bend right=10] node[near start, right] {$c_1$} (a);
			\path[line width = 0.8pt,->] (e) edge [bend right=10] node[below] {$c_1$} (a);							
		\end{tikzpicture} ,
	\end{align} and the second group 
	\begin{align}\label{wbasis_2}
		\begin{tikzpicture}[baseline={([yshift=-.5ex]current bounding box.center)}]
			\node[circle,draw] (a) at (6,2.7) {$w_1$};
			\node[circle,draw] (b) at (8.5,1.5) {$w_2$};		
			\node[circle,draw] (c) at (7.5,-.5) {$w_3$};
			\node[circle,minimum size =0.9cm, draw] (d) at (4.5,-.5) { };
			\node[circle,draw] (e) at (3.5,1.5) {$w_N$};	
			\path[every loop/.style={min distance=10mm,in=45,out=135,looseness=5,->},line width = 0.5pt] (a) edge [loop above]  node[above] {$c_1\oplus 1$}   (a);
			\path[every loop/.style={min distance=10mm,in=-45,out=45,looseness=5},line width = 0.5pt,->] (b) edge [loop above]  node[right] {$c_2$}  (b);
			\path[every loop/.style={min distance=10mm,in=270,out=0,looseness=5,->},line width = 0.5pt] (c) edge [loop above]  node[below] {$c_3$}  (c);
			\path[every loop/.style={min distance=10mm,in=185,out=265,looseness=5,->},line width = 0.5pt] (d) edge [loop above]   node[below] {$c_{N-1}$}  (d);
			\path[every loop/.style={min distance=10mm,in=135,out=-135,looseness=5,->},line width = 0.5pt] (e) edge [loop above]  node[left] {$c_N$}   (e);				
			\path[line width = 0.8pt,->] (a) edge[bend right=10]  node[above] {$c_N\oplus 1$}  (e);
			\path[line width = 0.8pt,->] (e) edge[bend right=10]  node[above] {$c_{N-1}$}  (d);
			\path[line width = 0.8pt,->] (c) edge[bend right=10]  node[above] {$c_2$}  (b);
			\path[line width = 0.8pt,->] (b) edge[bend right=10]  node[above] {$c_1$}  (a);
			\path[line width = 0.8pt,->] (a) edge  node[above] {$c_{N-1}\oplus 1$}  (d);
			\path[line width = 0.8pt,->] (a) edge  node[above] {$c_3\oplus 1$}  (c);
			\path[line width = 0.8pt,->] (a) edge[bend right=10]  node {$c_2\oplus 1$}  (b);
			\draw[thick, dashed] (5.2,-0.8) to[bend right=15] (6.8,-.8);		
			\draw (4.5,-.5) node {$w_{N-1}$};	
		\end{tikzpicture}.
	\end{align} 
	The above PM diagrams has $N$ PMs that correspond to the states of the W class. The $N=3$ W basis example is explained in Appendix~\ref{N=3_GHZW}.

	\paragraph{Dicke state generation} The Dicke state $|D_k^N\>$ ($1\le k \le N-1$) is a permutation symmetric state  with $k$ subsystems in $\downarrow$. Here, we provide a $\dc$ for the $k=2$ Dicke state.
	A $\dc$ for an arbitrary $k$ case can be conceived by induction from the W state case~\eqref{wstate_1} and the $D_2^N$ case that we will discuss here.
	The $k=2$ Dicke state can be expressed as
	\begin{align}
		|D_2^N\> = \frac{1}{\sqrt{\binom{N}{2}}} \sum_{p}|\downarrow_{p_1}\downarrow_{p_2},\uparrow_{p_3},\cdots,\uparrow_{p_N}\>
	\end{align} where the summation is over all the permutations of $(1,2,\cdots, N)$ that gives different states. This is represented as $\frac{N(N-1)}{2}$ PMs that consist of $(N-2)$ blue edges and $2$ red edges. 
	
	One can easily consider an LQN for $|D_2^N\>$ in analogy to the tritter LQN for the W state, i.e., sending $(N-2)$-particles in $\uparrow$ and $2$ particles in $\downarrow$ to $N$ detectors through a balanced unitary operator. For example, $|D_2^4\>$ can be generated by the following LQN:
	\begin{align}
		\begin{tikzpicture}[baseline={([yshift=-.5ex]current bounding box.center)}]
			\node[circle,draw] (1) at (0,0) {$w_3 $};
			\node[circle,draw] (2) at (2.5,0) {$w_4 $};
			\node[circle,draw] (3) at (2.5,2.5) {$w_2$};
			\node[circle,draw] (4) at (0,2.5) {$w_1$};
			\path[every loop/.style={min distance=10mm,in=-180,out=-90,looseness=5,},line width = 0.5pt,->,color=blue] (1) edge[loop above]   (1);
			\path[every loop/.style={min distance=10mm,in=-90,out=0,looseness=5},line width = 0.5pt,->,color=blue] (2) edge [loop above]   (2);
			\path[every loop/.style={min distance=10mm,in=0,out=90,looseness=5,color=red},line width = 0.5pt,->] (3) edge [loop above]   (3);	
			\path[every loop/.style={min distance=10mm,in=90,out=180,looseness=5,color=red},line width = 0.5pt,->] (4) edge [loop above]   (4);
			\path[line width = 0.8pt,->,color=blue] (1) edge[bend right=20]  node[near start] {$ $ } (2);
			\path[line width = 0.8pt,->,color=blue] (1) edge[bend right=20]  node[near start] {$ $ } (3);\path[line width = 0.8pt,->,color=blue] (1) edge[bend right=20]  node[near start] {$ $ } (4);
			\path[line width = 0.8pt,->,color=blue] (2) edge[bend right=20]  node[near start] {$ $ }  (1);
			\path[line width = 0.8pt,->,color=blue] (2) edge[bend right=20]  node[near start] {$ $ } (3);
			\path[line width = 0.8pt,->,color=blue] (2) edge[bend right=20]  node[near start] {$ $ } (4);
			\path[line width = 0.8pt,color=red,->] (3) edge[bend right=15]  node[near start] {$ $ } (2); 
			\path[line width = 0.8pt,color=red,->] (3) edge[bend right=15]  node[near start] {$ $ } (1);	
			\path[line width = 0.8pt,color=red,->] (3) edge[bend right=15]  node[near start] {$ $ } (4);
			\path[line width = 0.8pt,color=red,->] (4) edge[bend right=15]  node[near start] {$ $ } (1); 
			\path[line width = 0.8pt,color=red,->] (4) edge[bend right=15]  node[near start] {$ $ } (2);	
			\path[line width = 0.8pt,color=red,->] (4) edge[bend right=15]  node[near start] {$ $ } (3);
		\end{tikzpicture}.
	\end{align} 
	On the other hand, we can also generate $|D_2^N\>$ with an LQN of less edges in analogy to \eqref{wstate_1} for the W state.    
	We can conceive a $\dc$ for $|D_2^N\>$ based on the following $N$ loops
	\[\begin{tikzpicture}
		\node[circle,draw] (c) at (-2,0) {$w_3$};
		\node[circle,draw] (d) at (-0.5,0) {$w_4$};		
		\node[circle,draw] (e) at (2,0) {$w_{N}$};
		\node[circle,draw] (a) at (-1,1.5) {$w_{1}$};
		\node[circle,draw] (b) at (1,1.5) {$w_{2}$};
		\path[every loop/.style={min distance=10mm,in=45,out=135,looseness=5,color=red ,->},line width = 0.5pt] (a) edge [loop above]   (a);
		\path[every loop/.style={min distance=10mm,in=45,out=135,looseness=5,color=red ,->},line width = 0.5pt] (b) edge [loop above]   (b);
		\path[every loop/.style={min distance=10mm,in=-45,out=-135,looseness=5,color=blue ,->},line width = 0.5pt] (c) edge [loop above]  (c);
		\path[every loop/.style={min distance=10mm,in=-45,out=-135,looseness=5,color=blue ,->},line width = 0.5pt] (d) edge [loop above] (d);
		\path[every loop/.style={min distance=10mm,in=-45,out=-135,looseness=5,color=blue ,->},line width = 0.5pt] (e) edge [loop above]   (e);			
		\draw[thick, dotted] (0.3,0) -- (1.2,0) ;			
	\end{tikzpicture}\]	 
	and $2(N-2)$-cycles $(w_j\xrightarrow{R} w_k \xrightarrow{B} w_j)$ where $j=1,2$ and $k=3,4,\cdots ,N $:
	\begin{align}\label{D_2^N}
		\begin{tikzpicture}
			\node[circle,draw] (c) at (-3,0) {$w_3$};
			\node[circle,draw] (d) at (-1,0) {$w_4$};		
			\node[circle,draw] (e) at (3,0) {$w_{N}$};
			\node[circle,draw] (a) at (-1.2,2.5) {$w_{1}$};
			\node[circle,draw] (b) at (1.2,2.5) {$w_{2}$};
			\path[every loop/.style={min distance=10mm,in=45,out=135,looseness=5,color=red ,->},line width = 0.5pt] (a) edge [loop above]   (a);
			\path[every loop/.style={min distance=10mm,in=45,out=135,looseness=5,color=red ,->},line width = 0.5pt] (b) edge [loop above]   (b);
			\path[every loop/.style={min distance=10mm,in=-45,out=-135,looseness=5,color=blue ,->},line width = 0.5pt] (c) edge [loop above]  (c);
			\path[every loop/.style={min distance=10mm,in=-45,out=-135,looseness=5,color=blue ,->},line width = 0.5pt] (d) edge [loop above] (d);
			\path[every loop/.style={min distance=10mm,in=-45,out=-135,looseness=5,color=blue ,->},line width = 0.5pt] (e) edge [loop above]   (e);			
			\path[line width = 0.8pt,->,color=red] (a) edge [bend right=10] (c);		
			\path[line width = 0.8pt,->,color=red] (a) edge [bend right=15] (d);
			\path[line width = 0.8pt,->,color=red] (a) edge [bend right=10] (e);				
			\path[line width = 0.8pt,->,color=blue ] (c) edge [bend right=10] (a);	
			\path[line width = 0.8pt,->,color=blue ] (d) edge [bend right=15] (a);	
			\path[line width = 0.8pt,->,color=blue ] (e) edge [bend right=10] (a);
			\path[line width = 0.8pt,->,color=red] (b) edge [bend right=10] (c);		
			\path[line width = 0.8pt,->,color=red] (b) edge [bend right=15] (d);
			\path[line width = 0.8pt,->,color=red] (b) edge [bend right=10] (e);	
			\path[line width = 0.8pt,->,color=blue ] (c) edge [bend right=10] (b);	
			\path[line width = 0.8pt,->,color=blue ] (d) edge [bend right=15] (b);	
			\path[line width = 0.8pt,->,color=blue ] (e) edge [bend right=10] (b);		
			\draw[thick, dotted] (0,0) -- (1.5,0) ;			
		\end{tikzpicture}
	\end{align}
	Examining all elementary cycles, we see that the above LQN gives all PMs that construct $|D_2^N\>$ state (see Appendix~\ref{general_dicke} for  the final no-bunching state of the above $\dc$). However, unlike the LQNs for GHZ and W state we have proposed, not all PMs from \eqref{D_2^N} appear once. Indeed, by expressing a state with $i$th and $j$th subsystems in $\downarrow$ as $|\vec{n}_{ij}\>$, there are four states of $|\vec{n}_{kl}\>$ for $k\neq 1,2$ and $l\neq 1,2$. For the simplest $N=4$ example, \eqref{D_2^N} is drawn as
	\[\begin{tikzpicture}[baseline={([yshift=-.5ex]current bounding box.center)}]
		\node[circle,draw] (1) at (0,2.5) {$w_1 $};
		\node[circle,draw] (2) at (2.5,2.5) {$w_2 $};
		\node[circle,draw] (3) at (2.5,0) {$w_4$};
		\node[circle,draw] (4) at (0,0) {$w_3$};
		\path[every loop/.style={min distance=10mm,in=90,out=180,looseness=5,},line width = 0.5pt,->,color=red] (1) edge [loop above] (1);
		\path[every loop/.style={min distance=10mm,in=0,out=90,looseness=5},line width = 0.5pt,->,color=red] (2) edge [loop above]   (2);
		\path[every loop/.style={min distance=10mm,in=-90,out=0,looseness=5,color=blue},line width = 0.5pt,->] (3) edge [loop above]  (3);	
		\path[every loop/.style={min distance=10mm,in=-180,out=-90,looseness=5,color=blue},line width = 0.5pt,->] (4) edge [loop above]  (4);		
		
		\path[line width = 0.8pt,->,color=red] (1) edge[bend right=15]  node[near start] {$ $ } (3);
		\path[line width = 0.8pt,->,color=red] (1) edge[bend right=15]  node[near start] {$ $ } (4);

		\path[line width = 0.8pt,->,color=red] (2) edge[bend right=15]  node[near start] {$ $ } (3);
		\path[line width = 0.8pt,->,color=red] (2) edge[bend right=15]  node[near start] {$ $ } (4);
		\path[line width = 0.8pt,color=blue,->] (3) edge[bend right=15]  node[near start] {$ $ } (2); 
		\path[line width = 0.8pt,color=blue,->] (3) edge[bend right=15]  node[near start] {$ $ } (1);	
		
		\path[line width = 0.8pt,color=blue,->] (4) edge[bend right=15]  node[near start] {$ $ } (1); 
		\path[line width = 0.8pt,color=blue,->] (4) edge[bend right=15]  node[near start] {$ $ } (2);	
		
	\end{tikzpicture}.\]
	Its corresponding $G_{bb}$ and LQN are given in Fig.~\ref{linearq}.
	The postselected no-bunching state is given, with the  $T_{ij}$  as the amplitude from $i$ to $X_j$, by
	\begin{align}
		\begin{split}
			&T_{11}T_{22}T_{33}T_{44}|\downarrow_1\downarrow_2\uparrow_3\uparrow_4\> \\
			&+ T_{11}T_{44}T_{32}T_{23}|\downarrow_1\uparrow_2\downarrow_3\uparrow_4\> \\
			&+ T_{11}T_{33}T_{42}T_{24}|\downarrow_1\uparrow_2\uparrow_3\downarrow_4\>  \\
			& + T_{22}T_{44}T_{31}T_{13}|\uparrow_1\downarrow_2\downarrow_3\uparrow_4 \> \\
			&+ T_{14}T_{22}T_{33}T_{41}|\uparrow_1\downarrow_2\uparrow_3\downarrow_4 \>  \\
			&  +(T_{13}T_{31}T_{24}T_{42} +T_{23}T_{32}T_{14}T_{41} \\
			& \quad + T_{13}T_{32}T_{24}T_{41} + T_{14}T_{42}T_{23}T_{31})|\uparrow_1\uparrow_2\downarrow_3\downarrow_4\>.
		\end{split}
	\end{align} As we have mentioned, $|\uparrow_1\uparrow_2\downarrow_3\downarrow_4\> = |\vec{n}_{34}\>$ appears four times in the LQN. The amplitude of the state can be factorized as $(T_{13}T_{24}+T_{23}T_{14})(T_{31}T_{42} + T_{32}T_{41})|\uparrow_1\uparrow_2\downarrow_3\downarrow_4\>$.
	
	However, we can still achieve the standard $|D_2^4\>$ by properly fixing the amplitude phases. Indeed, it can be easily checked that the following imposition of amplitudes presents the standard $|D_2^4\>$:
	\begin{align}
		&T_{11} =T_{22}=T_{33}=T_{44}=\frac{1}{\sqrt{3}}, \nn \\
		&T_{13} = \frac{e^{\frac{i\pi}{6}}}{\sqrt{3}}=T_{31}^*,\quad  T_{14} = \frac{e^{\frac{-i\pi}{6}}}{\sqrt{3}}= T_{41}^*, \nn \\
		& T_{23} = \frac{e^{\frac{-i\pi}{6}}}{\sqrt{3}}= T_{32}^*,\quad T_{24} = \frac{e^{\frac{i\pi}{6}}}{\sqrt{3}}= T_{42}^*.
	\end{align}	
	
	We discusse further on the $N=5$ case in Appendix~\ref{general_dicke}.
	
	\section{Discussions}\label{discussion}
	By introducing a strict mapping relation of LQNs into graphs, we have shown that the entanglement generation in LQN can be efficiently analyzed with graph theory techniques. Other than the LQNs for genuine entanglement that we have explained here, we can find numerous LQNs for multipartite genuinely entangled states that have intriguing physical properties and can be verified experimentally~\footnote{S. Chin, D. Lee, and Y.-S. Kim, Multipartite entangled states of linear quantum networks: a graph theoretic approach, in preparation}. 
	
	At this point, it is worth emphasizing the distinctness of our graph picture for understanding entanglement in LQN from that in Refs.~\cite{krenn2017quantum,gu2019quantum,gu2019quantum2,gu2020quantum}. While both approaches exploit the graph theory for implementing experimental schemes to generate multipartite genuinely entangled states, the physical setups to be mapped to graphs are disparate. The graph analogies used in Refs.~\cite{krenn2017quantum,gu2019quantum,gu2019quantum2,gu2020quantum} describe multipartite entanglement generations with a mixture of probabilistic photon pair sources (spontaneous parametric down-conversion crystals, SPDC) and single-photon sources. Since most of the processes strongly depend on photon pair sources, we can consider that the analogies are based on optical systems. Considering the experimental demonstration, it is hardly scalable utilizing probabilistic photon pair sources. On the other hand, our graph picture analyzes multipartite entanglement generations with any kind of identical particles by linear transformations of spatial modes and  internal states. Therefore, it encompasses various multi-particle systems that evolve linearly, including fermionic systems as mentioned before (see, e.g., Eq.~\eqref{N=2_bipartite}). And this difference naturally result in the difference of the correspondence relations between two approaches. For example, vertices and (hyper-) edges correspond to output modes and photon sources in  Ref.~\cite{krenn2017quantum,gu2019quantum,gu2019quantum2,gu2020quantum}, which is fundamentally distinct from our correspondence relation in Table~\ref{dictionary}. 
		The reason why the final state is computed from perfect matchings in both works is that both postselect states without bunching at the output. However, since all the LQNs are mapped to only balanced bipartite graphs in our work, we can represent the graphs as corresponding directed graphs, which offers abundant mathematical benefit for analyzing the entanglement of identical particles.
		In addition, even when our scheme is realized with linear optical devices, it provides a practical advantage in scalability since it can utilize on-demand single photon sources~\cite{lee2021entangling}.
	
	We can apply our graph picture of LQNs to more general cases. We have focused so far on the cases where the particle number $N$ equals the mode number $M$ with the postselection of no-bunching state, i.e., $N=M$. And these systems are mapped to undirected balanced bipartite graphs that are isomorphic to perfect matching diagrams $\dc$. Our mapping relation can be directly generalized to any LQN cases with arbitrary $N$ and $M$. For example, the original boson sampling setup  with $N\ll M$ also postselects no-bunching states~\cite{aaronson2011computational}. And the relations in Table~\ref{dictionary} hold except that the elements of $V$ becomes $M$ instead of $N$ (unbalanced bipartite graphs). Thus, we expect that our computation technique for no-bunching states would be useful for understanding some features boson sampling. Moreover,
	we can also consider a mapping relation of LQNs and graphs when the final internal state $r_j^a$ of each particle can be a superposition of $\uparrow$ and $\downarrow$.
	Optics realizes this type of transformations with \emph{partially polarizing beamsplitters} (PPBS), which correspond to the partial collapse measurements that can suppress decoherence effects~\cite{kim2012protecting} and handily implement CNOT gates~\cite{langford2005demonstration,kiesel2005linear,okamoto2005demonstration}. While our current work have mapped LQNs into $G_{bb}$s that are simple (neither loops nor multiple edges), the generalized mapping would need multigraphs (permitting  multiple edges between vertices). One interesting application of such multigraph mapping would be cluster states~\cite{briegel2001persistent}, whose internal state basis defer among modes in general. The simplest case, i.e., $N=4$ example, is explained in Appendix~E, which we can achieve with simple graphs.

	\section*{Acknowledgements}
	We appreciate Dr. Hyang-Tag Lim, Dr. Young-Wook Cho, and Dongwha Lee for  helpful discussion. S.C. is also grateful to Prof. Jung-Hoon Chun for his support during the research. This work is supported by the National Research Foundation of Korea (NRF, NRF-2019R1I1A1A01059964 and NRF-2019M3E4A1079777). S.C is supported by the quantum computing technology development program of the National Research Foundation of Korea(NRF)funded by the Korean government (Ministry of Science and ICT(MSIT), No.2021M3H3A103657312).
	
	\appendix 

	\section{Graph theory glossary}\label{glossary}
	
	\paragraph{Graph---.} A \textbf{graph} $G=(V,E)$ is a collection of vertices $V$ and edges $E$ that connects the elements in $V$. Each edge can have  a \textbf{color} and a \textbf{weight}. An edge weight is a numerical value associated to the edge. In this paper, it is considered a complex number. 
	
	\paragraph{Undirected and directed graphs---.} Edges in a graph can have direction or not. An \textbf{undirected graph}  is made of undirected edges, and a \textbf{directed graph} (or digraph, $G_d$) of directed edges. For an undirected graph $G=(V,E)$ with $V= \{v_1,v_2,\cdots, v_N\}$, an edge ($\in E$) that connects $v_i$ and $v_j$ is denoted as $(v_i,v_j)$. 
	For a $G_d=(W,F)$ with $W=\{w_1,w_2,\cdots, w_N\}$, an edge ($\in F$) that connects $w_i$ and $w_j$ is denoted as $(w_i\to w_j)$.  
	
	For $G_d$, two vertices $w_i$ and $w_j$ are \textbf{strongly connected} if we can move from $w_i$ toward $w_j$ and from $w_j$ toward $w_i$ by  following the edge directions. If a path in a $G_d$ is of the form $(w_i\to \cdots \to w_i)$ and $w_i$ is the only repeated vertex in the path, it is a called an \textbf{elementary cycle}.     
	
	\paragraph{Bipartite graph and perfect matching---.}  A \textbf{bipartite graph} (or bigraph, $G_b=(U\cup V, E)$) has two disjoint sets of vertices, $U$ and $V$, such that every edge connects a vertex in $U$ to another vertex in $V$. A bigraph is \textbf{balanced} if the number of vertices in $U$ is equal to the number of vertices in $V$, i.e., $|U|=|V|$. Balanced bigraphs are denoted as $G_{bb}$.
	For example, with $U=\{a,b,c,d\}$ and $V=\{X,Y,Z,W\}$, a graph
	\begin{align}
		\begin{tikzpicture}
			\node[vertex,minimum size=0.5cm] (1) at (0,2) {$a$} ;
			\node[vertex,minimum size=0.5cm] (2) at (0,1) {$b$};
			\node[vertex,minimum size=0.5cm] (3) at (0,0) {$c$};
			\node[vertex,minimum size=0.5cm] (4) at (0,-1) {$d$};    
			\node[circle,draw,minimum size=0.8cm] (X) at (3,2) {$X$};
			\node[circle,draw,minimum size=0.8cm] (Y) at (3,1) {$Y$};
			\node[circle,draw,minimum size=0.8cm] (Z) at (3,0) {$Z$};	
			\node[circle,draw,minimum size=0.8cm] (W) at (3,-1) {$W$};	    
			\path[line width = 0.8pt] (1) edge (X);
			\path[line width = 0.8pt] (2) edge (Y);
			\path[line width = 0.8pt] (3) edge (Z);
			\path[line width = 0.8pt] (4) edge (W); 
			\path[line width = 0.8pt] (1) edge (Y);
			\path[line width = 0.8pt] (2) edge (Z);
			\path[line width = 0.8pt] (3) edge (W);
			\path[line width = 0.8pt] (4) edge (X);
		\end{tikzpicture} \nn \end{align}
	is a $G_{bb}$.
	
	A $G_{bb}$ can have \textbf{perfect matchings} (PMs). A PM is an independent set of edges in which every vertex of $U$ is connected to exactly one vertex of $V$. If an edge ($\in E$) is in a PM, it is \textbf{maximally matchable}.  
	There are two PMs of the above $G_{bb}$:
	\[\begin{tikzpicture}
		\node[vertex,minimum size=0.5cm] (1) at (0,2) {$a$} ;
		\node[vertex,minimum size=0.5cm] (2) at (0,1) {$b$};
		\node[vertex,minimum size=0.5cm] (3) at (0,0) {$c$};
		\node[vertex,minimum size=0.5cm] (4) at (0,-1) {$d$};    
		\node[circle,draw,minimum size=0.8cm] (X) at (2.5,2) {$X$};
		\node[circle,draw,minimum size=0.8cm] (Y) at (2.5,1) {$Y$};
		\node[circle,draw,minimum size=0.8cm] (Z) at (2.5,0) {$Z$};	
		\node[circle,draw,minimum size=0.8cm] (W) at (2.5,-1) {$W$};	    
		\path[line width = 0.8pt] (1) edge (X);
		\path[line width = 0.8pt] (2) edge (Y);
		\path[line width = 0.8pt] (3) edge (Z);
		\path[line width = 0.8pt] (4) edge (W); 
		
		\node[vertex,minimum size=0.5cm] (1) at (4,2) {$a$} ;
		\node[vertex,minimum size=0.5cm] (2) at (4,1) {$b$};
		\node[vertex,minimum size=0.5cm] (3) at (4,0) {$c$};
		\node[vertex,minimum size=0.5cm] (4) at (4,-1) {$d$};    
		\node[circle,draw,minimum size=0.8cm] (X) at (6.5,2) {$X$};
		\node[circle,draw,minimum size=0.8cm] (Y) at (6.5,1) {$Y$};
		\node[circle,draw,minimum size=0.8cm] (Z) at (6.5,0) {$Z$};	
		\node[circle,draw,minimum size=0.8cm] (W) at (6.5,-1) {$W$};	 
		\path[line width = 0.8pt] (1) edge (Y);
		\path[line width = 0.8pt] (2) edge (Z);
		\path[line width = 0.8pt] (3) edge (W);
		\path[line width = 0.8pt] (4) edge (X);
	\end{tikzpicture}\] 
	Since no edge is out of a PM, all edges of the $G_{bb}$ are maximally matchable.

	\paragraph{Adjacency matrix---.} The adjacency matrix of a graph is the matrix A with 
	\begin{align}
		\left\{ \begin{array}{ll}
			A_{ij}=0 & \textrm{if $v_i$ is not adjacent to $v_j$}\\
			A_{ij}\neq 0 & \textrm{if $v_i$ is adjacent to $v_j$}  
		\end{array} \right. \nn 
	\end{align} Since if $v_i$ is adjacent to $v_j$ then $v_j$ is adjacent to $v_i$ and vice versa in an undirected graph, the adjacency matrix of the graph is symmetric.  Conversely, if an adjacency matrix is symmetric, then we can regard its corresponding graph as undirected.

	By inserting the edge weights into the nonzero $A_{ij}$, we obtain the weighted adjacency matrix. By inserting the edge colors instead, we obtain the colored adjacency matrix. 
	
	\section{Linear transformation}	\label{lineartransformation}
	In this section, we show that the most general form of linear transformation can always be expressed as the simpler form of Eq.~\eqref{linear_single}. As in the main text, we set the initial state of the $a$th particle as $\P_a = (\p_a, s_a)$ ($\p_a$: the particle location, $s_a$: the internal state of dimension $d$), and the computational basis at the detector level as $\Phi_j = (\phi_j, r_j)$ ($\phi_j$: the detector location, $r_j$: the internal state). 
	Then, the linearity relation between $\P_a$ and $\Phi_j$ is expressed in the most general form as 
	\begin{align}\label{linear_generalist}
		|\P_a\>=	|(\p_a,s_a)\> &\to  \sum_{j=1}^N\sum_{r_j=\uparrow,\downarrow} T_{as_a,jr_j}|(\phi_j, r_j)\>, \nn \\
		&\qquad \qquad \quad  (T_{as_a,jr_j} \in \mathbb{C})  
	\end{align} where the complex number $T_{as_a,jr_j}$ is normalized as $\sum_{j,r_j}|T_{as_a,jr_j}|^2 =1$.  However, since $s_a$ is fixed for each $a$ as an input, $T_{as_a,jr_j}$ can be simply denoted without loss of generality as $T_{a,jr_j}$. Then Eq.~\eqref{linear_generalist} is rewritten as 
	\begin{align}\label{linear_general}
		|\P_a\>&\to \sum_{j=1}^N\sum_{r_j=\uparrow,\downarrow} T_{a,jr_j}|(\phi_j, r_j)\> \nn \\
		&\sum_{j=1}^N (T_{a,j\uparrow}|(\phi_j, \uparrow)\> +  T_{a,j\downarrow}|(\phi_j, \downarrow)\>).\nn \\
		&\qquad (T_{a,jr_j} \in \mathbb{C},\quad \sum_{j,r_j}|T_{a,jr_j}|^2 =1)  
	\end{align} 
	Furthermore, since we are interested in the tranformed internal state according to the path from $\p_a$ to $\phi_j$, $T_{a,jr_j}$ can be rewritten as
	\begin{align}\label{T_devision}
		T_{a,jr_j} = T_{aj}\a_{ajr_j}, \quad (T_{aj}, \a_{ajr_j} \in \mathbb{C})
	\end{align} with $\sum_{j}|T_{aj}|^2 =1$ and $\sum_{r_j}|\a_{a,jr_j}|^2=1$ (note that $|T_{a,j\uparrow}|^2 + |T_{a,j\downarrow}|^2 = |T_{aj}|^2$ holds with these conditions).
	Substituting Eq.~\eqref{T_devision} into Eq.~\eqref{linear_general}, we have
	\begin{align}\label{linear_simpler}
		|\P_a\>\to \sum_{j=1}^N T_{a,j}(\a_{aj\uparrow}|(\phi_j, \uparrow)\> +  \a_{aj\downarrow}|(\phi_j, \downarrow)\>)
	\end{align} 
	By setting the internal state as 
	\begin{align}
		\a_{aj\uparrow}|\uparrow\> + \a_{aj\downarrow}|\downarrow\> \equiv |r_{j}^a\>, 	
	\end{align}
	We obtain Eq.~\eqref{linear_single}. 
	
	\section{$N=3$ GHZ and W basis}\label{N=3_GHZW}
	\paragraph{$N=3$ GHZ basis.---} For the $N=3$ case, the $G_{bb}$-representation of \eqref{GHZ_class} becomes
	\begin{align}
		\begin{tikzpicture}[baseline={([yshift=-.5ex]current bounding box.center)}]
			\node[vertex] (1) at (0,3.2) {$1$} ;
			\node[vertex] (2) at (0,1.6) {$2$};
			\node[vertex] (3) at (0,0) {$3$};
			\node[circle,draw,minimum size=0.8cm] (X) at (4,3.2) {$X_1$};
			\node[circle,draw,minimum size=0.8cm] (Y) at (4,1.6) {$X_2$};
			\node[circle,draw,minimum size=0.8cm] (Z) at (4,0) {$X_3$};	
			\path[line width = 0.8pt] (1) edge node [near start] {$c_1$} (X);
			\path[line width = 0.8pt] (2) edge  node [near start] {$c_2$} (Y);
			\path[line width = 0.8pt] (3) edge  node [near start] {$c_3$} (Z);
			\path[line width = 0.8pt] (1) edge node {$c_2\oplus 1$} (Y);
			\path[line width = 0.8pt] (2) edge node [near end] {$c_3\oplus 1$} (Z);
			\path[line width = 0.8pt] (3) edge node [near start] {$c_1\oplus 1$} (X);
		\end{tikzpicture},
		\nn \end{align}
	which has two PMs
	\begin{align}
		&
		\begin{tikzpicture}[baseline={([yshift=-.5ex]current bounding box.center)}]
			\node[vertex] (1) at (0,3) {$1$} ;
			\node[vertex] (2) at (0,1.5) {$2$};
			\node[vertex] (3) at (0,0) {$3$};
			\node[circle,draw,minimum size=0.8cm] (X) at (2,3) {$X_1$};
			\node[circle,draw,minimum size=0.8cm] (Y) at (2,1.5) {$X_2$};
			\node[circle,draw,minimum size=0.8cm] (Z) at (2,0) {$X_3$};	
			\path[line width = 0.8pt] (1) edge node  {$c_1$} (X);
			\path[line width = 0.8pt] (2) edge  node  {$c_2$} (Y);
			\path[line width = 0.8pt] (3) edge  node {$c_3$} (Z);
		\end{tikzpicture} \quad+\quad 
		\begin{tikzpicture}[baseline={([yshift=-.5ex]current bounding box.center)}]
			\node[vertex] (1) at (0,3) {$1$} ;
			\node[vertex] (2) at (0,1.5) {$2$};
			\node[vertex] (3) at (0,0) {$3$};
			\node[circle,draw,minimum size=0.8cm] (X) at (2,3) {$X_1$};
			\node[circle,draw,minimum size=0.8cm] (Y) at (2,1.5) {$X_2$};
			\node[circle,draw,minimum size=0.8cm] (Z) at (2,0) {$X_3$};	
			\path[line width = 0.8pt] (1) edge node {$c_2\oplus 1$} (Y);
			\path[line width = 0.8pt] (2) edge node[near end] {$c_3\oplus 1$} (Z);
			\path[line width = 0.8pt] (3) edge node {$c_1\oplus 1$} (X);
		\end{tikzpicture}  \nn \\
		& =T_{11}T_{22}T_{33}|c_1, c_2, c_3\>  \nn \\
		&\quad + T_{12}T_{23}T_{31}|c_1\oplus 1, c_2\oplus 2,c_3\oplus 3\>.
	\end{align} By varying $(c_1,c_2,c_3)$, we obtain the eight states that construct a GHZ basis (see, e.g.,~\cite{cunha2019tripartite}).
	
	\paragraph{$N=3$ W basis.---}
	From \eqref{wbasis_1}, we have
	\begin{align}
		&\begin{tikzpicture}
			\node[circle,draw] (1) at (0,2.2) {$w_1$};
			\node[circle,draw] (2) at (1.5,0) {$w_2$};
			\node[circle,draw] (3) at (-1.5,0) {$w_3$};
			\path[every loop/.style={min distance=10mm,in=45,out=135,looseness=5},line width = 0.5pt,->,color=black] (1) edge [loop above]  node[above] {$c_1\oplus 1$} (1);
			\path[every loop/.style={min distance=10mm,in=270,out=0,looseness=5},line width = 0.5pt,->,color=black] (2) edge [loop above]  node[below] {$c_2$} (2);
			\path[every loop/.style={min distance=10mm,in=180,out=270,looseness=5},line width = 0.5pt, color=black,->] (3) edge [loop above]  node[below] {$c_3$} (3);
			\path[line width = 0.8pt,->,color=black] (1) edge[bend left=15]  node[right] {$c_2\oplus 1$} (2);
			\path[line width = 0.8pt,color=black,->] (2) edge[bend left=15]  node[below] {$c_1$} (1);
			\path[line width = 0.8pt,color=black,->] (1) edge[bend left=15]  node[below] {$c_3\oplus 1$} (3);    
			\path[line width = 0.8pt,->,color=black] (3) edge[bend left=15]  node[left] {$c_1$} (1);      
		\end{tikzpicture} \nn
	\end{align} whose $G_{bb}$ is
	\begin{align} 
		\begin{tikzpicture}[baseline={([yshift=-.5ex]current bounding box.center)}]
			\node[vertex] (1) at (0,3.2) {$1$} ;
			\node[vertex] (2) at (0,1.6) {$2$};
			\node[vertex] (3) at (0,0) {$3$};
			\node[circle,draw,minimum size=0.8cm] (X) at (4,3.2) {$X_1$};
			\node[circle,draw,minimum size=0.8cm] (Y) at (4,1.6) {$X_2$};
			\node[circle,draw,minimum size=0.8cm] (Z) at (4,0) {$X_3$};	
			\path[line width = 0.8pt] (1) edge node [near start] {$c_1\oplus 1$} (X);
			\path[line width = 0.8pt] (2) edge  node [near start] {$c_2$} (Y);
			\path[line width = 0.8pt] (3) edge  node [near start] {$c_3$} (Z);
			\path[line width = 0.8pt] (1) edge node [near start] {$c_2\oplus 1$} (Y);
			\path[line width = 0.8pt] (2) edge node [near end] {$c_1$} (X);
			\path[line width = 0.8pt] (3) edge node [near start] {$c_1$} (X);
			\path[line width = 0.8pt] (1) edge node [near end] {$c_3\oplus 1 $} (Z);
		\end{tikzpicture}.
	\end{align} There are three PMs,
	\begin{align}\label{wbasis_final}
		&
		\begin{tikzpicture}[baseline={([yshift=-.5ex]current bounding box.center)}]
			\node[vertex] (1) at (0,3) {$1$} ;
			\node[vertex] (2) at (0,1.5) {$2$};
			\node[vertex] (3) at (0,0) {$3$};
			\node[circle,draw,minimum size=0.8cm] (X) at (2,3) {$X_1$};
			\node[circle,draw,minimum size=0.8cm] (Y) at (2,1.5) {$X_2$};
			\node[circle,draw,minimum size=0.8cm] (Z) at (2,0) {$X_3$};	
			\path[line width = 0.8pt] (1) edge node  {$c_1\oplus 1$} (X);
			\path[line width = 0.8pt] (2) edge  node  {$c_2$} (Y);
			\path[line width = 0.8pt] (3) edge  node {$c_3$} (Z);
		\end{tikzpicture} \quad+\quad 
		\begin{tikzpicture}[baseline={([yshift=-.5ex]current bounding box.center)}]
			\node[vertex] (1) at (0,3) {$1$} ;
			\node[vertex] (2) at (0,1.5) {$2$};
			\node[vertex] (3) at (0,0) {$3$};
			\node[circle,draw,minimum size=0.8cm] (X) at (2,3) {$X_1$};
			\node[circle,draw,minimum size=0.8cm] (Y) at (2,1.5) {$X_2$};
			\node[circle,draw,minimum size=0.8cm] (Z) at (2,0) {$X_3$};	
			\path[line width = 0.8pt] (1) edge node[near start] {$c_2 \oplus 1$} (Y);
			\path[line width = 0.8pt] (2) edge node[near start] {$c_1$} (X);
			\path[line width = 0.8pt] (3) edge node {$c_3$} (Z);
		\end{tikzpicture} \nn \\
		& + \quad 
		\begin{tikzpicture}[baseline={([yshift=-.5ex]current bounding box.center)}]
			\node[vertex] (1) at (0,3) {$1$} ;
			\node[vertex] (2) at (0,1.5) {$2$};
			\node[vertex] (3) at (0,0) {$3$};
			\node[circle,draw,minimum size=0.8cm] (X) at (2,3) {$X_1$};
			\node[circle,draw,minimum size=0.8cm] (Y) at (2,1.5) {$X_2$};
			\node[circle,draw,minimum size=0.8cm] (Z) at (2,0) {$X_3$};	
			\path[line width = 0.8pt] (1) edge node [near start] {$c_3 \oplus 1$} (Z);
			\path[line width = 0.8pt] (2) edge  node [near start] {$c_2$} (Y);
			\path[line width = 0.8pt] (3) edge  node [near start] {$c_1$} (X);
		\end{tikzpicture} \nn \\
		= & T_{11}T_{22}T_{33}|c_1\oplus 1,c_2,c_3\> \nn \\
		& + T_{21}T_{12}T_{33}|c_1, c_2\oplus 1,c_3\> \nn\\
		& +T_{31}T_{13}T_{22}|c_1, c_2,c_3\oplus 1\>, 
	\end{align} which construct the W basis~\cite{cunha2019tripartite}.
	
	From \eqref{wbasis_2}, we have
	\begin{align}
		&\begin{tikzpicture}[baseline={([yshift=-.5ex]current bounding box.center)}]
			\node[circle,draw,minimum size=0.6cm] (1) at (0,2.2) {$w_1$};
			\node[circle,draw,minimum size=0.6cm] (2) at (1.5,0) {$ w_2$};
			\node[circle,draw,minimum size=0.6cm] (3) at (-1.5,0) {$w_3 $};
			\path[every loop/.style={min distance=10mm,in=45,out=135,looseness=5},line width = 0.5pt,->,color=black] (1) edge [loop above]  node[above] {$c_1\oplus 1$} (1);
			\path[every loop/.style={min distance=10mm,in=270,out=0,looseness=5},line width = 0.5pt,->,color=black] (2) edge [loop above]  node[below] {$c_2$} (2);
			\path[every loop/.style={min distance=10mm,in=180,out=270,looseness=5},line width = 0.5pt, color=black,->] (3) edge [loop above]  node[below] {$c_3$}  (3);
			\path[line width = 0.8pt,->,color=black] (1) edge[bend left=15]  node[right] {$c_2\oplus 1$} (2);
			\path[line width = 0.8pt,color=black,->] (2) edge[bend left=15] node[below] {$c_1$} (1);
			\path[line width = 0.8pt,color=black,->] (3) edge  node[below] {$c_2$} (2);
			\path[line width = 0.8pt,color=black,->] (1) edge  node[left] {$c_3\oplus 1$} (3);    
		\end{tikzpicture},
		\nn
	\end{align} whose $G_{bb}$ is 
	\begin{align} 
		\begin{tikzpicture}[baseline={([yshift=-.5ex]current bounding box.center)}]
			\node[vertex,minimum size=0.5cm] (1) at (0,3.2) {$1$} ;
			\node[vertex,minimum size=0.5cm] (2) at (0,1.6) {$2$};
			\node[vertex,minimum size=0.5cm] (3) at (0,0) {$3$};
			\node[circle,draw,minimum size=0.8cm] (X) at (4,3.2) {$X_1$};
			\node[circle,draw,minimum size=0.8cm] (Y) at (4,1.6) {$X_2$};
			\node[circle,draw,minimum size=0.8cm] (Z) at (4,0) {$X_3$};	
			\path[line width = 0.8pt] (1) edge node [near start] {$c_1\oplus 1$} (X);
			\path[line width = 0.8pt] (2) edge  node [near start] {$c_2$} (Y);
			\path[line width = 0.8pt] (3) edge  node [near start] {$c_3$} (Z);
			\path[line width = 0.8pt] (1) edge node [near start] {$c_2\oplus 1$} (Y);
			\path[line width = 0.8pt] (2) edge node [near end] {$c_1$} (X);
			\path[line width = 0.8pt] (3) edge node [near start] {$c_2$} (Y);
			\path[line width = 0.8pt] (1) edge node [near start] {$c_3\oplus 1 $} (Z);
		\end{tikzpicture},
	\end{align} which has three PMs, 
	\begin{align}&
		\begin{tikzpicture}[baseline={([yshift=-.5ex]current bounding box.center)}]
			\node[vertex,minimum size=0.5cm] (1) at (0,3) {$1$} ;
			\node[vertex,minimum size=0.5cm] (2) at (0,1.5) {$2$};
			\node[vertex,minimum size=0.5cm] (3) at (0,0) {$3$};
			\node[circle,draw,minimum size=0.8cm] (X) at (2,3) {$X_1$};
			\node[circle,draw,minimum size=0.8cm] (Y) at (2,1.5) {$X_2$};
			\node[circle,draw,minimum size=0.8cm] (Z) at (2,0) {$X_3$};	
			\path[line width = 0.8pt] (1) edge node  {$c_1\oplus 1$} (X);
			\path[line width = 0.8pt] (2) edge  node  {$c_2$} (Y);
			\path[line width = 0.8pt] (3) edge  node {$c_3$} (Z);
		\end{tikzpicture} \quad+\quad 
		\begin{tikzpicture}[baseline={([yshift=-.5ex]current bounding box.center)}]
			\node[vertex,minimum size=0.5cm] (1) at (0,3) {$1$} ;
			\node[vertex,minimum size=0.5cm] (2) at (0,1.5) {$2$};
			\node[vertex,minimum size=0.5cm] (3) at (0,0) {$3$};
			\node[circle,draw,minimum size=0.8cm] (X) at (2,3) {$X_1$};
			\node[circle,draw,minimum size=0.8cm] (Y) at (2,1.5) {$X_2$};
			\node[circle,draw,minimum size=0.8cm] (Z) at (2,0) {$X_3$};	
			\path[line width = 0.8pt] (1) edge node[near start] {$c_2 \oplus 1$} (Y);
			\path[line width = 0.8pt] (2) edge node[near start] {$c_1$} (X);
			\path[line width = 0.8pt] (3) edge node {$c_3$} (Z);
		\end{tikzpicture} \nn \\
		& + \quad 
		\begin{tikzpicture}[baseline={([yshift=-.5ex]current bounding box.center)}]
			\node[vertex,minimum size=0.5cm] (1) at (0,3) {$1$} ;
			\node[vertex,minimum size=0.5cm] (2) at (0,1.5) {$2$};
			\node[vertex,minimum size=0.5cm] (3) at (0,0) {$3$};
			\node[circle,draw,minimum size=0.8cm] (X) at (2,3) {$X_1$};
			\node[circle,draw,minimum size=0.8cm] (Y) at (2,1.5) {$X_2$};
			\node[circle,draw,minimum size=0.8cm] (Z) at (2,0) {$X_3$};	
			\path[line width = 0.8pt] (1) edge node {$c_3 \oplus 1$} (Z);
			\path[line width = 0.8pt] (2) edge  node[near end] {$c_1$} (X);
			\path[line width = 0.8pt] (3) edge  node {$c_2$} (Y);
		\end{tikzpicture} \nn \\
		= & T_{11}T_{22}T_{33}|c_1\oplus 1,c_2,c_3\> \nn \\
		& + T_{21}T_{12}T_{33}|c_1, c_2\oplus 1,c_3\> \nn\\
		& +T_{21}T_{32}T_{13}|c_1, c_2,c_3\oplus 1\>. 
	\end{align} We see that the final result is equal to \eqref{wbasis_final}.
	
	\section{$D_2^N$ state generation}\label{general_dicke}	
	
	The postselected no-bunching state from \eqref{D_2^N} is given by
	\begin{widetext}
		\begin{align}
			&\big(\prod_{i=1}^{N}T_{ii}\big)|\vn_{12}\> + \sum_{j=3}^{N}\big(\prod_{\substack{i\neq 2,j\\ =1} }^{N}T_{ii}\big)T_{j2}T_{2j}|\vn_{1j}\> + \sum_{j=3}^{N}\big(\prod_{\substack{i\neq j\\=2}}^{N}T_{ii}\big)T_{j1}T_{1j}|\vn_{2j}\>  \nn \\
			&+ \sum_{3\le j< k\le N}	\big( \prod_{\substack{i\neq j,k\\ =3}}^{N}T_{ii}\big)(T_{1j}T_{2k}+T_{2j}T_{1k})(T_{j2}T_{k1}+T_{j1}T_{k2})|\vn_{jk}\>,
		\end{align}
		where $|\vec{n}_{ij}\>$ denotes a state with $i$th and $j$th subsystems in $\downarrow$.
		One can see that there are four states of $|\vec{n}_{kl}\>$ for $k\neq 1,2$ and $l\neq 1,2$.
		We can check whether the standard $D_2^N$ state from the above state can be obtained by controlling the amplitudes.
		For the $N=5$ case, the $\dc$ is drawn as
		\[\begin{tikzpicture}
			\node[circle,draw] (c) at (-3,0) {$w_3$};
			\node[circle,draw] (d) at (0,0) {$w_4$};		
			\node[circle,draw] (e) at (3,0) {$w_{5}$};
			\node[circle,draw] (a) at (-1.5,2) {$w_{1}$};
			\node[circle,draw] (b) at (1.5,2) {$w_{2}$};
			\path[every loop/.style={min distance=10mm,in=45,out=135,looseness=5,color=red ,->},line width = 0.5pt] (a) edge [loop above]   (a);
			\path[every loop/.style={min distance=10mm,in=45,out=135,looseness=5,color=red ,->},line width = 0.5pt] (b) edge [loop above]   (b);
			\path[every loop/.style={min distance=10mm,in=-45,out=-135,looseness=5,color=blue ,->},line width = 0.5pt] (c) edge [loop above]  (c);
			\path[every loop/.style={min distance=10mm,in=-45,out=-135,looseness=5,color=blue ,->},line width = 0.5pt] (d) edge [loop above] (d);
			\path[every loop/.style={min distance=10mm,in=-45,out=-135,looseness=5,color=blue ,->},line width = 0.5pt] (e) edge [loop above]   (e);			
			\path[line width = 0.8pt,->,color=red] (a) edge [bend left=10] (c);		
			\path[line width = 0.8pt,->,color=red] (a) edge [bend left=15] (d);
			\path[line width = 0.8pt,->,color=red] (a) edge [bend left=10] (e);				
			\path[line width = 0.8pt,->,color=blue ] (c) edge [bend left=10] (a);	
			\path[line width = 0.8pt,->,color=blue ] (d) edge [bend left=15] (a);	
			\path[line width = 0.8pt,->,color=blue ] (e) edge [bend left=10] (a);
			\path[line width = 0.8pt,->,color=red] (b) edge [bend left=10] (c);		
			\path[line width = 0.8pt,->,color=red] (b) edge [bend left=15] (d);
			\path[line width = 0.8pt,->,color=red] (b) edge [bend left=10] (e);	
			\path[line width = 0.8pt,->,color=blue ] (c) edge [bend left=10] (b);	
			\path[line width = 0.8pt,->,color=blue ] (d) edge [bend left=15] (b);	
			\path[line width = 0.8pt,->,color=blue ] (e) edge [bend left=10] (b);		
		\end{tikzpicture}.\]	
		Then, the final no-bunching state is given by
		\begin{align}
			\begin{split}
				& T_{11}T_{22}T_{33}T_{44}T_{55}|\downarrow_1\downarrow_2\uparrow_3\uparrow_4\uparrow_5\>+
				T_{11}T_{44}T_{55}T_{32}T_{23}|\downarrow_1\uparrow_2\downarrow_3\uparrow_4\uparrow_5\> + T_{11}T_{33}T_{55}T_{42}T_{24}|\downarrow_1\uparrow_2\uparrow_3\downarrow_4\uparrow_5\> \\ &+T_{11}T_{33}T_{44}T_{52}T_{25}|\downarrow_1\uparrow_2\uparrow_3\uparrow_4\downarrow_5\> + T_{22}T_{44}T_{55}T_{31}T_{13}|\uparrow_1\downarrow_2\downarrow_3\uparrow_4\uparrow_5\> + T_{22}T_{33}T_{55}T_{14}T_{41}|\uparrow_1\downarrow_2\uparrow_3\downarrow_4\uparrow_5\> \\
				&+  T_{22}T_{33}T_{44}T_{15}T_{51}|\uparrow_1\downarrow_2\uparrow_3\uparrow_4\downarrow_5\> +T_{55}(T_{13}T_{24}+T_{23}T_{14})(T_{32}T_{41}+T_{31}T_{42})|\uparrow_1\uparrow_2\downarrow_3\downarrow_4\uparrow_5\> \\
				&+T_{44}(T_{13}T_{25}+T_{23}T_{15})(T_{32}T_{51}+T_{31}T_{52})|\uparrow_1\uparrow_2\downarrow_3\uparrow_4\downarrow_5\>
				\\
				&+T_{33}(T_{14}T_{25}+T_{24}T_{15})(T_{42}T_{51}+T_{41}T_{52})|\uparrow_1\uparrow_2\uparrow_3\downarrow_4\downarrow_5\>.
			\end{split}
		\end{align}

		We obtain the standard $D_2^5$ state by setting the transformation amplitudes as
		\begin{align}
			\begin{split}
				&T_{11} =T_{22}=\frac{1}{\sqrt{4}},\quad T_{33}=T_{44}=T_{55}=\frac{1}{\sqrt{3}}, 
				\\
				&T_{13} = \frac{1}{\sqrt{4}},\quad T_{14} =\frac{e^{-\frac{i\pi}{3}}}{\sqrt{4}},\quad T_{15} =\frac{e^{-\frac{2i\pi}{3}}}{\sqrt{4}},  \qquad (T_{j1} =T_{1j}^*) 
				\\
				& T_{23} = \frac{e^{-\frac{i\pi}{3} } }{\sqrt{4}}, \quad  T_{24} = \frac{1}{\sqrt{4}},\quad  T_{25} = \frac{e^{\frac{i\pi}{3} } }{\sqrt{4}}.
				\qquad (T_{j2} =T_{2j}^*) 
			\end{split}
		\end{align}
		
	\end{widetext}	
	
	\section{$N=4$ cluster state generation}\label{cluster}
	The $N=4$ cluster state is expressed as
	\begin{align}
		|C_4\> = \frac{1}{2}\big(&|\uparrow_1\uparrow_2\uparrow_3\uparrow_4\> + |\uparrow_1\uparrow_2\downarrow_3\downarrow_4\> \nn \\
		& +|\downarrow_1\downarrow_2\uparrow_3\uparrow_4\> -|\downarrow_1\downarrow_2\downarrow_3\downarrow_4\> \big).
	\end{align}
	
	To build an LQN that carries the above state, we need  to draw a bipartite graph with four PMs, i.e.,\\ $ $ \\
	\begin{tikzpicture}[baseline={([yshift=-.5ex]current bounding box.center)}]
		\node[vertex] (1) at (0,2) {1} ;
		\node[vertex] (2) at (0,1) {2};
		\node[vertex] (3) at (0,0) {3};
		\node[vertex] (4) at (0,-1) {4};    
		\node[circle,draw,minimum size=0.8cm] (X) at (2.5,2) {$X$};
		\node[circle,draw,minimum size=0.8cm] (Y) at (2.5,1) {$Y$};
		\node[circle,draw,minimum size=0.8cm] (Z) at (2.5,0) {$Z$};	
		\node[circle,draw,minimum size=0.8cm] (W) at (2.5,-1) {$W$};	    
		\path[line width = 0.8pt,color=blue] (1) edge (X);
		\path[line width = 0.8pt,color=blue] (2) edge (Y);
		\path[line width = 0.8pt,color=blue] (3) edge (Z);
		\path[line width = 0.8pt,color=blue] (4) edge (W); 
		
		\node[vertex] (1) at (4,2) {1} ;
		\node[vertex] (2) at (4,1) {2};
		\node[vertex] (3) at (4,0) {3};
		\node[vertex] (4) at (4,-1) {4};    
		\node[circle,draw,minimum size=0.8cm] (X) at (6.5,2) {$X$};
		\node[circle,draw,minimum size=0.8cm] (Y) at (6.5,1) {$Y$};
		\node[circle,draw,minimum size=0.8cm] (Z) at (6.5,0) {$Z$};	
		\node[circle,draw,minimum size=0.8cm] (W) at (6.5,-1) {$W$};	    
		\path[line width = 0.8pt,color=red] (1) edge (X);
		\path[line width = 0.8pt,color=red] (2) edge (Y);
		\path[line width = 0.8pt,color=red] (3) edge (Z);
		\path[line width = 0.8pt,color=red] (4) edge (W); 
	\end{tikzpicture} \\
	$ $ \\
	$ $ \\
	$ $ \\
	\begin{tikzpicture}[baseline={([yshift=-.5ex]current bounding box.center)}]
		\node[vertex] (1) at (0,2) {1} ;
		\node[vertex] (2) at (0,1) {2};
		\node[vertex] (3) at (0,0) {3};
		\node[vertex] (4) at (0,-1) {4};    
		\node[circle,draw,minimum size=0.8cm] (X) at (2.5,2) {$X$};
		\node[circle,draw,minimum size=0.8cm] (Y) at (2.5,1) {$Y$};
		\node[circle,draw,minimum size=0.8cm] (Z) at (2.5,0) {$Z$};	
		\node[circle,draw,minimum size=0.8cm] (W) at (2.5,-1) {$W$};	    
		\path[line width = 0.8pt,color=red] (1) edge (X);
		\path[line width = 0.8pt,color=red] (2) edge (Y);
		\path[line width = 0.8pt,color=blue] (3) edge (Z);
		\path[line width = 0.8pt,color=blue] (4) edge (W);

		\node[vertex] (1) at (4,2) {1} ;
		\node[vertex] (2) at (4,1) {2};
		\node[vertex] (3) at (4,0) {3};
		\node[vertex] (4) at (4,-1) {4};    
		\node[circle,draw,minimum size=0.8cm] (X) at (6.5,2) {$X$};
		\node[circle,draw,minimum size=0.8cm] (Y) at (6.5,1) {$Y$};
		\node[circle,draw,minimum size=0.8cm] (Z) at (6.5,0) {$Z$};	
		\node[circle,draw,minimum size=0.8cm] (W) at (6.5,-1) {$W$};	    
		\path[line width = 0.8pt,color=blue] (1) edge (X);
		\path[line width = 0.8pt,color=blue] (2) edge (Y);
		\path[line width = 0.8pt,color=red] (3) edge (Z);
		\path[line width = 0.8pt,color=red] (4) edge (W); 
	\end{tikzpicture}.
	$ $ \\
	
	Note that this state can be seen as a superposition of two GHZ states of different local bases,
	\[
	|\uparrow_1\uparrow_2\uparrow_3\uparrow_4\> + |\downarrow_1\downarrow_2\downarrow_3\downarrow_4\>,~~ 
	|\uparrow_1\uparrow_2\downarrow_3\downarrow_4\> -|\downarrow_1\downarrow_2\uparrow_3\uparrow_4\>.
	\]
	Therefore, we can use the $\dc$ for $N=4$ GHZ state~\eqref{GHZ_PM} for our current goal. By adding two red edges to the $N=4$ GHZ $\dc$, we obtain a $\dc$ of the following form:    
	\[\begin{tikzpicture}[baseline={([yshift=-.5ex]current bounding box.center)}]
		\node[circle,draw] (1) at (0,2.5) {$w_1 $};
		\node[circle,draw] (2) at (2.5,2.5) {$w_2 $};
		\node[circle,draw] (3) at (2.5,0) {$w_3$};
		\node[circle,draw] (4) at (0,0) {$w_4$};
		\path[every loop/.style={min distance=10mm,in=90,out=180,looseness=5,},line width = 0.5pt,->,color=blue] (1) edge [loop above] (1);
		\path[every loop/.style={min distance=10mm,in=0,out=90,looseness=5},line width = 0.5pt,->,color=blue] (2) edge [loop above]   (2);
		\path[every loop/.style={min distance=10mm,in=-90,out=0,looseness=5,color=blue},line width = 0.5pt,->] (3) edge [loop above]  (3);	
		\path[every loop/.style={min distance=10mm,in=-180,out=-90,looseness=5,color=blue},line width = 0.5pt,->] (4) edge [loop above]  (4);		
		\path[line width = 0.8pt,->,color=red] (1) edge[bend right=15]  node[near start] {$ $ } (2);
		\path[line width = 0.8pt,->,color=red] (2) edge[bend right=15]  node[near start] {$ $ } (1);
		
		\path[line width = 0.8pt,->,color=red] (3) edge[bend right=15]  node[near start] {$ $ } (4);
		\path[line width = 0.8pt,->,color=red] (4) edge[bend right=15]  node[near start] {$ $ } (3);			
		
		\path[line width = 0.8pt,->,color=red] (2) edge  node[near start] {$ $ } (3);
		
		\path[line width = 0.8pt,color=red,->] (4) edge node[near start] {$ $ } (1); 
		
	\end{tikzpicture}.\]
	The above $\dc$  satisfies the two conditions of Theorem~\ref{genuine_ent} and has three elementary cycles. The above $\dc$ is drawn in the $G_{bb}$ representation as	
	\[
	\begin{tikzpicture}[baseline={([yshift=-.5ex]current bounding box.center)}]
		\node[vertex] (1) at (0,2) {1} ;
		\node[vertex] (2) at (0,1) {2};
		\node[vertex] (3) at (0,0) {3};
		\node[vertex] (4) at (0,-1) {4};    
		\node[circle,draw,minimum size=0.8cm] (X) at (2.5,2) {$X$};
		\node[circle,draw,minimum size=0.8cm] (Y) at (2.5,1) {$Y$};
		\node[circle,draw,minimum size=0.8cm] (Z) at (2.5,0) {$Z$};	
		\node[circle,draw,minimum size=0.8cm] (W) at (2.5,-1) {$W$};	    
		\path[line width = 0.8pt,color=blue] (1) edge (X);
		\path[line width = 0.8pt,color=blue] (2) edge (Y);
		\path[line width = 0.8pt,color=blue] (3) edge (Z);
		\path[line width = 0.8pt,color=blue] (4) edge (W); 
		\path[line width = 0.8pt,color=red] (4) edge (X);
		\path[line width = 0.8pt,color=red] (1) edge (Y);
		\path[line width = 0.8pt,color=red] (2) edge (Z);
		\path[line width = 0.8pt,color=red] (3) edge (W); 
		\path[line width = 0.8pt,color=red] (2) edge (X);
		\path[line width = 0.8pt,color=red] (4) edge (Z);
	\end{tikzpicture}, \]
	which has four PMs that can construct $|C_4\>$ with a LQN. \\
	$ $\\
		$ $\\
			$ $\\
				$ $\\
					$ $\\

\bibliographystyle{unsrtdin}
\bibliography{GL}

\end{document}